\documentclass[prd,preprint,tightenlines,floatfix,showpacs,preprintnumbers,nofootinbib,eqsecnum]{revtex4}

 \usepackage[dvips,final]{graphicx}
  \usepackage{amssymb}
   \usepackage{amsmath}
    \usepackage{amsfonts}
     \usepackage{epsfig}
      \usepackage{bm}

      \usepackage{multirow}

\newcommand{\GeV}{\textrm{GeV}}

\begin{document}

\thispagestyle{empty} \preprint{\hbox{}} \vspace*{-10mm}

\title{Nonperturbative and spin effects \\ in the central exclusive production
of tensor $\chi_c(2^+)$ meson}

\author{R.~S.~Pasechnik}
\email{roman.pasechnik@fysast.uu.se} \affiliation{High Energy
Physics, Department of Physics and Astronomy, Uppsala University Box
535, SE-75121 Uppsala, Sweden}

\author{A.~Szczurek}
\email{antoni.szczurek@ifj.edu.pl} \affiliation{Institute of Nuclear
Physics PAN, PL-31-342 Cracow,
Poland and\\
University of Rzesz\'ow, PL-35-959 Rzesz\'ow, Poland}

\author{O.~V.~Teryaev}
\email{teryaev@theor.jinr.ru} \affiliation{Bogoliubov Laboratory of
Theoretical Physics, JINR, Dubna 141980, Russia}

\date{\today}

\begin{abstract}
We discuss central exclusive production (CEP) of the tensor
$\chi_c(2^{+})$ meson in proton-(anti)proton collisions at Tevatron,
RHIC and LHC energies. The amplitude for the process is derived
within the $k_t$-factorisation approach. Differential and total
cross sections are calculated for several unintegrated gluon
distributions (UGDFs). We compare exclusive production of all
charmonium states $\chi_c(0^+),\,\chi_c(1^+)$ and $\chi_c(2^+)$.
Equally good description of the recent Tevatron data is achieved
both with Martin-Ryskin phenomenological UGDF and UGDF based on
unified BFKL-DGLAP approach. Unlike for Higgs production, the main
contribution to the diffractive amplitude of heavy quarkonia comes
from nonperturbative region of gluon transverse momenta
$Q_{\perp}<1\,\GeV$. At $y \approx$ 0, depending on UGDF we predict
the contribution of $\chi_c(1^+,2^+)$ to the $J/\Psi + \gamma$
channel to be comparable or larger than that of the $\chi_c(0^+)$
one. This is partially due to a significant contribution from lower
polarization states $\lambda=0$ for $\chi_c(1^+)$ and
$\lambda=0,\,\pm 1$ for $\chi_c(2^+)$ meson. Corresponding
theoretical uncertainties are discussed.
\end{abstract}


\maketitle

\section{Introduction}

\begin{figure}[!h]    
 \centerline{\includegraphics[width=0.35\textwidth]{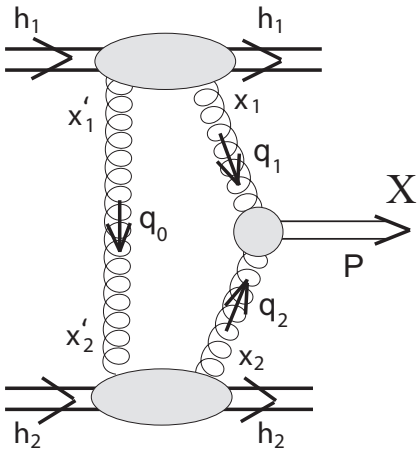}}
   \caption{\label{fig:fig1}
   \small \em  The QCD mechanism of diffractive production of
   the heavy central system $X$.}
\end{figure}

It is well known that the exclusive diffractive Higgs production
provides a very convenient tool for Higgs searches at hadron
colliders due to a very clean environment unlike the inclusive
production \cite{KMR}. A QCD mechanism for the diffractive
production of heavy central system has been proposed by
Kaidalov, Khoze, Martin and Ryskin (Durham group) for Higgs
production at the LHC (see Refs.~\cite{KMR,KKMR,KKMR-spin}). Below
we will refer to it as the KKMR approach. In the framework of this
approach the amplitude of the exclusive $pp\to pXp$ process is
considered to be a convolution of the hard subprocess amplitude
describing fusion of two off-shell gluons producing a heavy system
$g^*g^*\to X$, and the soft hadronic factors containing information
about emission of the relatively soft gluons from the proton lines
(see Fig.~\ref{fig:fig1}). In the framework of
the $k_{\perp}$-factorisation approach these soft parts are written in
terms of so-called off-diagonal unintegrated gluon distributions
(UGDFs). The QCD factorisation is rigorously justified in the limit
of very large factorisation scale being the transverse mass of
the central system $M_{\perp}$.

In order to check the underlying production mechanism it is worth to
replace the Higgs boson by a lighter (but still heavy enough to provide
the QCD factorisation) meson which is easier to measure. In this
respect the exclusive production of heavy quarkonia is under special
interest from both experimental and theoretical point of view
\cite{exp-thr}. Verifying the KKMR approach against various data on
exclusive meson production at high energies is a good test of
nonperturbative dynamics of parton distributions encoded in UGDFs.

Recently, the signal from the diffractive $\chi_c(0^+,1^+,2^+)$
charmonia production in the radiative $J/\Psi + \gamma$ decay
channel has been measured by the CDF Collaboration
\cite{Aaltonen:2009kg}: $d\sigma/dy|_{y=0}(pp\to
pp(J/\psi+\gamma))\simeq(0.97\pm0.26)$ nb. Assuming the absolute
dominance of the spin-0 contribution, this result was published by
the CDF Collaboration in the form:
\[\frac{d\sigma}{dy}\Big|_{y=0}(\chi_c(0^+))\simeq
\frac{1}{\mathrm{BR}(\chi_c(0^+)\to
J/\psi+\gamma)}\frac{d\sigma}{dy}\Big|_{y=0}(pp\to
pp(J/\psi+\gamma))=(76\pm14)\,\mathrm{nb}.\]
Indeed, in the very forward limit the contributions from
$\chi_c(1^+,2^+)$ vanish due to the $J_z=0$ selection rule
\cite{Khoze:2004yb,Khoze:2000jm}. This is not true, however, for
general kinematics \cite{Pasechnik:2007hm,Martin:2009ku}. In
particular, it was shown in Ref.~\cite{Pasechnik:2009bq} that the
axial-vector $\chi_c(1^+)$ production, due to a relatively large
branching fraction of its radiative decay, may not be negligible and
gives a significant contribution to the total signal measured by the
CDF Collaboration. The same holds also for the tensor $\chi_c(2^+)$
meson contribution \cite{ourpos}. Recent Durham group
investigations \cite{HarlandLang:2009qe} support these predictions.

The production of the axial-vector $\chi_c(1^+)$ meson is
additionally suppressed w.r.t. $\chi_c(0^+,2^+)$ in the limit of
on-shell fusing gluons (with non-forward protons) due to the
Landau-Yang theorem \cite{Pasechnik:2009bq}. Such an extra
suppression may, in principle, lead to the dominance of the
$\chi_c(2^+)$ contribution over the $\chi_c(1^+)$ one in the
radiative decay channel \cite{ourpos}. Off-shell effects play a
significant role even for the scalar $\chi_c(0^+)$ production
reducing the total cross section by a factor of 2 -- 5 depending on
UGDFs \cite{Pasechnik:2007hm}.
The major part of the amplitude comes from rather small
gluon transverse momenta $Q_{\perp}<1\,\GeV$. This requires a
special attention and including all polarisation states
$\chi_c(1^+,2^+)$.
Our present goal is to analyze these
issues in more detail in the case of tensor charmonium production at
the Tevatron, to study its energy dependence and to compare with
corresponding contributions from scalar and axial-vector charmonia.

The paper is organized as follows. Section 2 contains the
generalities of the QCD central exclusive production mechanism, two
different prescriptions for off-diagonal UGDFs are introduced and
discussed. In Section 3 we derive the hard subprocess amplitude
$g^*g^*\to\chi_c(2^+)$ in the nonrelativistic QCD formalism and
consider its properties. Section 4 contains numerical results for
total and differential cross sections of $\chi_c(0^+,1^+,2^+)$ CEP
and their correspondence to the last CDF data. In Section 5 the
summary of main results is given.

\section{Diffractive $pp\to pp\chi_c(2^+)$ production amplitude}

The general kinematics of the central exclusive production (CEP)
process $pp\to pXp$ with $X$ being the colour singlet $q{\bar q}$
bound state has already been discussed in our previous papers on
$\chi_c(0^+)$ \cite{Pasechnik:2007hm} and $\chi_c(1^+)$
\cite{Pasechnik:2009bq} production. In this section we adopt the
same notations and consider the matrix element for exclusive
$\chi_c(2^+)$ production and its properties in detail.

According to the KKMR approach the amplitude of the exclusive double
diffractive color singlet production $pp\to pp\chi_{cJ}$ is
\cite{Khoze:2004yb,Pasechnik:2007hm}
\begin{eqnarray}
{\cal M}^{p p \to p p \chi_{cJ}}_{J,\lambda}=s\cdot
\pi^2\frac12\frac{\delta_{c_1c_2}}{N_c^2-1}\,\int d^2
q_{0,t}V^{c_1c_2}_{J,\lambda}(q_1,q_2,p_M) \;
\nonumber \\
\times\frac{f^{\mathrm{off}}_{g,1}(x_1,x_1',q_{0,t}^2,q_{1,t}^2,t_1)
      f^{\mathrm{off}}_{g,2}(x_2,x_2',q_{0,t}^2,q_{2,t}^2,t_2)}
{q_{0,t}^2\,q_{1,t}^2\, q_{2,t}^2} \; ,\label{ampl}
\end{eqnarray}
where $t_{1,2}$ are the momentum transfers along the proton lines,
$q_0$ is the momentum of the screening gluon, $q_{1,2}$ are the
momenta of fusing gluons, and
$f^{\mathrm{off}}_{g,i}(x_i,x_i',q_{0,t}^2,q_{i,t}^2,t_i)$ are the
off-diagonal UGDFs (see Fig.~\ref{fig:fig1}).

Traditional (asymmetric) form of the off-diagonal UGDFs is taken in
the limit of very small $x'\ll x_{1,2}$ in analogy to collinear
off-diagonal gluon distributions (with factorized $t$-dependence)
\cite{Kimber:2001sc,MR}, i.e.
\begin{eqnarray}
f_{g,1}^{\mathrm{off}} &=& R_g\,
f_g^{(1)}(x_1,Q^{\mathrm{eff}\,{}^2}_{1,t},\mu^2) \cdot F_N(t_1)\, ,
\nonumber \\
f_{g,2}^{\mathrm{off}} &=& R_g\,
f_g^{(2)}(x_2,Q^{\mathrm{eff}\,{}^2}_{2,t},\mu^2) \cdot
F_N(t_2),\quad \mu^2=\frac{M_{\perp}^2}{4} \label{asym-off}
\end{eqnarray}
with a quasiconstant prefactor $R_g$ which accounts for the single
$\log Q^2$ skewed effect \cite{Shuvaev:1999ce} and is found to be
$1.4$ at the Tevatron energy and $1.2$ at the LHC energy (for LO
PDF),
$Q^{\mathrm{eff}\,{}^2}_{1/2,t}=\mathrm{min}(q_{0,t}^2,q_{1/2,t}^2)$
are the effective gluon transverse momenta, as adopted in
Ref.~\cite{KMR,Khoze:2004yb}, $F_N(t)$ is the proton vertex factor,
which can be parameterized as $F_N(t)=\exp(b_0t)$ with
$b_0=2\,\GeV^{-2}$ \cite{KMR00}, or by the isoscalar nucleon form
factor $F_1(t)$ as we have done in Ref.~\cite{Pasechnik:2007hm}.
Below we shall refer to Eq.~(\ref{asym-off}) as KMR UGDF\footnote{In
actual calculations we use a more precise phenomenological
Martin-Ryskin UGDF introduced in Ref.~\cite{MR}. We are very
thankful to L.~Harland-Lang for a discussion on this point.}.

Our results in Ref.~\cite{Pasechnik:2007hm} showed up a strong
sensitivity of the KMRS numerical results \cite{Khoze:2004yb} on the
definition of the effective gluon transverse momenta
$Q^{\mathrm{eff}}_{1/2,t}$ and the factorisation scales $\mu_{1,2}$.
This behavior is explained by the fact that for $\chi_c$ production
the great part of the diffractive amplitude (\ref{ampl}) comes from
nonperturbatively small $q_{0,t}<1\,\GeV$. It means that the total
diffractive process is dominated by very soft screening gluon
exchanges with no hard scale and extremely small $x'\ll x_{1,2}$.

In principle, the factor $R_g$ in Eq.~(\ref{asym-off}) should be a
function of $x'$ and $x_1$ or $x_2$. In this case the off-diagonal
UGDFs do not depend on $x'$ and $q_{0,t}^2$ (or $q_{1/2,t}^2$), and
their evolution is reduced to diagonal UGDFs evolution corresponding
to one ``effective'' gluon. In general, the factor $R_g$ can depend on
UGDF and reflects complicated and still not well known dynamics in
the small-$x$ region.

In order to test this small-$x$ dynamics and estimate the
theoretical uncertainties related to introducing one ``effective''
gluon transverse momentum instead of two ones in
Eq.~(\ref{asym-off}), in Refs.~\cite{Pasechnik:2007hm,SPT07} we have
used more generalized symmetrical prescription for the off-diagonal
UGDFs. Actually, it is possible to calculate the off-diagonal UGDFs
in terms of their diagonal counterparts as follows\footnote{For
diagonal distributions without explicit scale dependences
the $\mu_0^2$, $\mu^2$ arguments must be omitted.}
\begin{eqnarray}\nonumber
f_{g,1}^{\mathrm{off}} &=& \sqrt{f_{g}^{(1)}(x_1',q_{0,t}^2,\mu_0^2)
\cdot
f_{g}^{(1)}(x_1,q_{1,t}^2,\mu^2)} \cdot F_N(t_1)\,, \\
f_{g,2}^{\mathrm{off}} &=& \sqrt{f_{g}^{(2)}(x_2',q_{0,t}^2,\mu_0^2)
\cdot f_{g}^{(2)}(x_2,q_{2,t}^2,\mu^2)} \cdot F_N(t_2)\,,
\label{skewed-UGDFs}
\end{eqnarray}
where
\begin{eqnarray}\nonumber
x'_1=x'_2,\quad \mu_0^2=q_{0,t}^2,\quad \mu^2=\frac{M_{\perp}^2}{4}.
\end{eqnarray}
This form of skewed two-gluon UGDFs (\ref{skewed-UGDFs}) is inspired
by the positivity constraints for the collinear Generalized Parton
Distributions \cite{posit}, and can be considered as a saturation of
the Cauchy-Schwarz inequality for the density matrix
\cite{PhysRept}. It allows us to incorporate the actual dependence
of the off-diagonal UGDFs on longitudinal momentum fraction of the
soft screening gluon $x'$ and its transverse momentum $q_{0,t}^2$ in
explicitly symmetric way. As will be shown below, these
symmetric off-diagonal UGDFs lead to results which are consistent
with the Tevatron data.

However, trying to incorporate the actual dependence of UGDFs on
(small but nevertheless finite) $x'$ we may encounter a problem. The
kinematics of the double diffractive process $pp\to pXp$ does not
give any precise expression for $x'$ in terms of the phase space
integration variables. From the QCD mechanism under consideration
one can only expect the general inequality $x'\ll x_{1,2}$ and upper
bound $x'\lesssim q_{0,t}/\sqrt{s}$ since the only scale appearing
in the left part of the gluon ladder is the transverse momentum of
the soft screening gluon $q_{0,t}$.

To explore the sensitivity of the final results on the values of
$x'$, staying in the framework of traditional KKMR approach, one can
introduce naively $x'=\xi\cdot q_{0,t}/\sqrt{s}$ with an auxiliary
parameter $\xi$ \cite{Pasechnik:2009bq}. In our earlier papers
\cite{Pasechnik:2007hm,SPT07} we considered the limiting case of
maximal $x'$ (with $\xi=1$). However, it is worth to compare the
predictions of the underlying QCD mechanism for smaller $\xi$
against the available experimental data in order to estimate typical
$x'$ values. We will analyze this issue in greater detail in the
Results section.

\section{Hard subprocess $g^*g^*\to\chi_c(2^+)$ amplitude}

Projection of the hard amplitude onto the singlet charmonium bound
state $V_{\mu\nu}^{c_{1}c_{2}}$ is given by an 4-dimentional
integral over relative momentum of quark and antiquark
$q=(k_1-k_2)/2$ \cite{HKSST00,HKSST01}:
\begin{eqnarray}\nonumber
&&V_{J,\,\mu\nu}^{c_{1}c_{2}}(k_1,k_2)={\cal
P}(q\bar{q}\rightarrow\chi_{cJ})\bullet
\Psi^{c_1c_2}_{ik,\,\mu\nu}(k_1,k_2)=
2\pi\cdot\sum_{i,k}\sum_{L_{z},S_{z}}\frac{1}{\sqrt{m}}\int
\frac{d^{\,4}q}{(2\pi )^{4}}\delta \left(
q^{0}-\frac{{\bf q}^{2}}{M}\right)\times\\
&&\times\,\Phi_{L=1,L_{z}}({\bf q})\cdot\left\langle
L=1,L_{z};S=1,S_{z}|J,J_{z}\right\rangle \left\langle
3i,\bar{3}k|1\right\rangle {\rm
Tr}\left\{\Psi_{ik,\,\mu\nu}^{c_{1}c_{2}}{\cal
P}_{S=1,S_{z}}\right\}, \label{amplitude-diff} \\
&&\Psi_{ik,\,\mu\nu}^{c_{1}c_{2}}=-g^2\biggl[t^{c_1}_{ij}t^{c_2}_{jk}\cdot
\biggl\{\gamma_{\nu}\frac{\hat{q}_{1,t}-\hat{k}_{1,t}-m}
{(q_1-k_1)^2-m^2}\gamma_{\mu}\biggr\}-t^{c_2}_{kj}t^{c_1}_{ji}\cdot
\biggl\{\gamma_{\mu}\frac{\hat{q}_{1,t}-\hat{k}_{2,t}+m}{(q_1-k_2)^2-m^2}
\gamma_{\nu}\biggr\}\biggr].\nonumber
\end{eqnarray}
Here the function $\Phi_{L=1,L_{z}}({\bf q})$ is the momentum space
wave function of the charmonium, the Clebsch-Gordan coefficient in
color space is
$\left\langle3i,\bar{3}k|1\right\rangle=\delta^{ik}/\sqrt{N_{c}},$
the trace of $t$-matrices is ${\rm
Tr}\,(t^{c_1}t^{c_2})=\delta^{c_1c_2}/2$, and the projection
operator ${\cal P}_{S=1,S_{z}}$ for a small relative momentum $q$
has the form
\begin{eqnarray} \label{P}
{\cal
P}_{S=1,S_{z}}=\frac{1}{2m}(\hat{k}_2-m)\frac{\hat{\epsilon}(S_{z})}
{\sqrt{2}}(\hat{k}_1+m).
\end{eqnarray}

Since $P$-wave function $\Phi_{L=1,L_{z}}$ vanishes at the origin,
we may expand the trace in Eq.~(\ref{amplitude-diff}) in the Taylor
series around ${\bf q}=0$, and only the linear terms in $q^{\sigma}$
survive. This yields an expression proportional to
\begin{eqnarray}\label{expansion}
\int \frac{d^{3}{\bf q}}{(2\pi)^{3}}q^{\sigma}\Phi_{L=1,L_{z}}({\bf
q})=-i \sqrt{\frac{3}{4\pi}}\epsilon^{\sigma}(L_{z}){\cal
R}^{\prime}(0),
\end{eqnarray}
with the derivative of the $P$-wave radial wave function at the
origin ${\cal R}^{\prime}(0)$ whose numerical value can be found in
Ref.~\cite{EQ95}. The general $P$-wave result (\ref{amplitude-diff})
may be further reduced by employing the Clebsch-Gordan identity
which for the tensor $\chi_{cJ=2}$ charmonium states reads
\begin{eqnarray*}
{\cal T}^{\sigma\rho}_{J=2}\equiv\sum_{L_{z},S_{z}}\!\!\left\langle
1,L_{z};1,S_{z}|2,J_{z}\right\rangle\epsilon^{\sigma}(L_{z})
\epsilon^{\rho}(S_{z})=\epsilon^{\sigma\rho}(J_{z}).
\end{eqnarray*}

Taking into account standard definitions of the light-cone vectors
$n^+=p_2/E_{cms},\; n^-=p_1/E_{cms}$ and momentum decompositions
$q_1=x_1p_1+q_{1,t},\; q_2=x_2p_2+q_{2,t}$ and using the gauge
invariance property (Gribov's trick) one gets the following
projection (for any spin $J$)
\begin{eqnarray}
&&\quad q_1^{\nu}V^{c_1c_2}_{J,\,\mu\nu}=
q_2^{\mu}V^{c_1c_2}_{J,\,\mu\nu}=0, \nonumber\\
V^{c_1c_2}_{J}(q_1,q_2)&=&
n^+_{\mu}n^-_{\nu}V_{J,\,\mu\nu}^{c_1c_2}(q_1,q_2)=
\frac{4}{s}\frac{q^{\nu}_{1,t}}{x_1}\frac{q^{\mu}_{2,t}}{x_2}
V^{c_1c_2}_{J,\,\mu\nu}(q_1,q_2). \label{decomp}
\end{eqnarray}
Since we adopt here the definition of the polarization vectors
proportional to gluon transverse momenta $q_{1/2,t}$, then
\begin{equation}
V_{J,\lambda}^{c_1,c_2}(q_{1,t},q_{2,t}) \to 0,\quad q_{1,t} \to
0,\quad\mathrm{or}\quad q_{2,t} \to 0\,.
\end{equation}
It shows that gluon transverse momenta are necessary to get a
nonzeroth diffractive cross section.

Summarizing all ingredients above, we get the vertex factor
$g^*g^*\to\chi_c(2^+)$ in the following covariant form
\begin{eqnarray}\nonumber
&&V^{c_1c_2}_{J=2}=2ig^2\sqrt{\frac{3}{M\pi N_c}}
\frac{\delta^{c_1c_2}{\cal
R}'(0)\epsilon_{\rho\sigma}^{(\lambda)}}{MM_{\perp}^2(q_1q_2)^2}\,
\biggl[(q_{1,t}q_{2,t})(q_1^{\sigma}-q_2^{\sigma})
\Big\{P^{\rho}(q_{1,t}^2-q_{2,t}^2)+(x_1p_1^{\rho}-x_2p_2^{\rho})M^2-\\
&&(q_{1,t}^{\rho}-q_{2,t}^{\rho})M^2\Big\}-2(q_1q_2)\Big\{M^2(q_{1,t}^{\rho}q_{2,t}^{\sigma}+q_{1,t}^{\sigma}q_{2,t}^{\rho})-
q_{1,t}^2(q_{1,t}^{\rho}q_{2,t}^{\sigma}+q_{2,t}^{\sigma}q_{2,t}^{\rho})-\label{Vgen-chi2}\\
\nonumber &&
q_{2,t}^2(q_{1,t}^{\sigma}q_{2,t}^{\rho}+q_{1,t}^{\sigma}q_{1,t}^{\rho})+
(x_1p_1^{\sigma}-x_2p_2^{\sigma})(q_{1,t}^2q_{2,t}^{\rho}-q_{2,t}^2q_{1,t}^{\rho})+
(q_{1,t}q_{2,t})(x_1p_1^{\rho}-x_2p_2^{\rho})(q_{1,t}^{\sigma}-q_{2,t}^{\sigma})-\\
\nonumber &&
2q_{1,t}^2x_1p_1^{\rho}q_{2,t}^{\sigma}-2q_{2,t}^2x_2p_2^{\rho}q_{1,t}^{\sigma}+
2(q_{1,t}q_{2,t})(x_1p_1^{\sigma}q_{2,t}^{\rho}+x_2p_2^{\sigma}q_{1,t}^{\rho})+
\frac{M_{\perp}^2}{s}(q_{1,t}q_{2,t})(p_1^{\rho}p_2^{\sigma}+p_2^{\rho}p_1^{\sigma})\Big\}
\end{eqnarray}

Polarization tensor of $\chi_{cJ=2}$ satisfies the following
relations (see e.g. Ref.~\cite{FillionGourdeau:2007ee})
\begin{eqnarray*}
&&P^{\mu}\epsilon_{\mu\nu}(\lambda)=P^{\nu}\epsilon_{\mu\nu}(\lambda)=0,\qquad
\epsilon_{\mu\nu}(\lambda)=\epsilon_{\nu\mu}(\lambda),\qquad
\epsilon_{\mu\mu}(\lambda)=0,\qquad
\epsilon_{\mu\nu}(\lambda){\epsilon^{\mu\nu}}^*(\lambda')=\delta_{\lambda\lambda'},\\
&&\sum_{\lambda=0,\pm1,\pm2}\epsilon_{\mu\nu}(\lambda){\epsilon_{\rho\sigma}}^*(\lambda)=
\frac12M_{\mu\rho}M_{\nu\sigma}+\frac12M_{\mu\sigma}M_{\nu\rho}-\frac13M_{\mu\nu}M_{\rho\sigma},\qquad
M_{\mu\nu}=g_{\mu\nu}-\frac{P_{\mu}P_{\nu}}{M^2}
\end{eqnarray*}
One can check that it may be represented in the following general
form
\begin{eqnarray}\label{poltens}
\epsilon_{\mu\nu}(\pm|\lambda|)&=&\frac{\sqrt{6}}{2}\delta_{0|\lambda|}
\left(n_3^{\mu}n_3^{\nu}+\frac{1}{3}\left[g_{\mu\nu}-\frac{P_{\mu}P_{\nu}}{M^2}\right]\right)\\
\nonumber
&&+\frac12\delta_{1|\lambda|}\Big(i[n_2^{\mu}n_3^{\nu}+n_3^{\mu}n_2^{\nu}]
\pm[n_1^{\mu}n_3^{\nu}+n_3^{\mu}n_1^{\nu}]\Big)
\\
\nonumber
&&-\frac12\delta_{2|\lambda|}\Big(i[n_1^{\mu}n_2^{\nu}+n_2^{\mu}n_1^{\nu}]
\pm[n_1^{\mu}n_1^{\nu}-n_2^{\mu}n_2^{\nu}]\Big) \; ,
\end{eqnarray}
where $n_{1,2,3}$ are light-like basis vectors satisfying
$n^{\mu}_{\alpha}n^{\nu}_{\beta}g_{\mu\nu}=g_{\alpha\beta}$ (with
$n^{\mu}_0=P_{\mu}/M$), and $\lambda=0,\pm1,\pm2$ are the
$\chi_c(2^+)$ meson helicities. To our best knowledge, there is no
explicit decomposition of the meson polarisation tensor
$\epsilon_{\mu\nu}(\lambda)$ in terms of basis vectors $n_i$ like
Eq.~(\ref{poltens}) in the literature. In practical calculations
below it is convenient to use it in a different representation:
\begin{eqnarray*}
\epsilon_{\mu\nu}(\lambda)&=&\frac{\sqrt{6}}{12}(2-|\lambda|)(1-|\lambda|)\left[g_{\mu\nu}-\frac{P_{\mu}P_{\nu}}{M^2}\right]+
\frac{\sqrt{6}}{4}(2-|\lambda|)(1-|\lambda|)n_3^{\mu}n_3^{\nu}+\\
\nonumber
&&+\frac14\lambda(1-|\lambda|)[n_1^{\mu}n_1^{\nu}-n_2^{\mu}n_2^{\nu}]+
\frac14i|\lambda|(1-|\lambda|)[n_1^{\mu}n_2^{\nu}+n_2^{\mu}n_1^{\nu}]+\\
\nonumber
&&+\frac12\lambda(2-|\lambda|)[n_1^{\mu}n_3^{\nu}+n_3^{\mu}n_1^{\nu}]+
\frac12i|\lambda|(2-|\lambda|)[n_2^{\mu}n_3^{\nu}+n_3^{\mu}n_2^{\nu}]
\; .
\end{eqnarray*}

Similarly to what has been done for $\chi_c(1^+)$ production in
Ref.~\cite{Pasechnik:2009bq}, in the c.m.s. frame we choose the
basis with collinear ${\bf n}_3$ and ${\bf P}$ vectors (so, we have
${\bf P}=(E,0,0,P_z),\;P_z=|{\bf P}|>0$) as a simplest one
\begin{eqnarray}
n_1^{\beta}=(0,\,1,\,0,\,0),\;\; n_2^{\beta}=(0,\,0,\,1,\,0),\;\;
n_3^{\beta}=\frac{1}{M}\,(|{\bf P}|,0,0,E),\;\; |{\bf
P}|=\sqrt{E^2-M^2}. \label{basis}
\end{eqnarray}
%
\begin{figure}[h!]
 \centerline{\epsfig{file=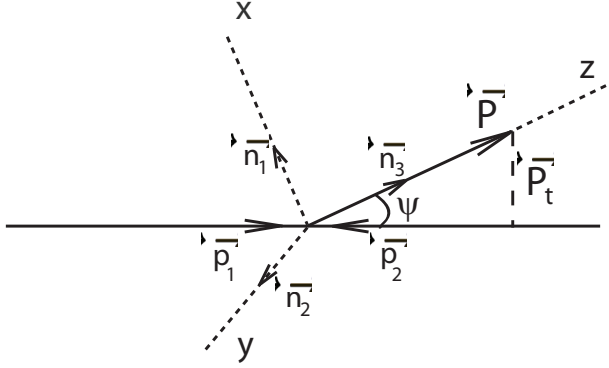,height=5cm,width=8cm}}
 \caption{\small \em Coordinate basis in the center-of-mass system
 of incoming protons $p_{1,2}$.}
 \label{fig:cms}  \end{figure}
Note, that we choose ${\bf n}_2$ to be transverse to the c.m.s beam
axis (see Fig.~\ref{fig:cms}), while ${\bf n}_1,\,{\bf n}_3$ are
turned around by the polar angle $\psi=[0\,...\,\pi]$ between ${\bf
P}$ and the c.m.s. beam axis. In the considered basis $\{{\bf
n}_1,\,{\bf n}_2,\,{\bf n}_3\}$ we have the following coordinates of
the incoming protons
\begin{eqnarray}\label{protons}
p_1=\frac{\sqrt{s}}{2}(1,\,-\sin\psi,\,0,\,\cos\psi),\quad
p_2=\frac{\sqrt{s}}{2}(1,\,\sin\psi,\,0,\,-\cos\psi) \; .
\end{eqnarray}

The gluon transverse momenta with respect to the c.m.s. beam axis
are
\begin{eqnarray*}\nonumber
q_{1,t}=(0,\,Q_{1,t}^{x}\cos\psi,\,Q_{t}^y,\,Q_{1,t}^{x}\sin\psi),\quad
q_{2,t}=(0,\,Q_{2,t}^{x}\cos\psi,\,-Q_{t}^y,\,Q_{2,t}^{x}\sin\psi),
\end{eqnarray*}
where $Q_{1/2,t}^{x},\,\pm Q_{t}^{y}$ are the components of the
gluon transverse momenta in the basis with the $z$-axis collinear to
the c.m.s. beam axis.

From definition (\ref{protons}) it follows that energy of the meson
and polar angle $\psi$ are related to covariant scalar products in
the considered coordinate system as \cite{Pasechnik:2009bq}
\begin{eqnarray}\label{Epsi}
E=\frac{(p_1P)+(p_2P)}{\sqrt{s}},\quad
\cos\psi=\frac{(p_1P)-(p_2P)}{\sqrt{s}|{\bf P}|},\quad
\sin\psi=\frac{(p_2n_1)-(p_1n_1)}{\sqrt{s}}.
\end{eqnarray}
Furthermore, we also see that from $q_1=x_1p_1+q_{1,t},\;
q_2=x_2p_2+q_{2,t}$ and $q_1+q_2=P$ we have
\begin{eqnarray}\label{x1x2cos}
x_1=\frac{E+|{\bf P}|\cos\psi}{\sqrt{s}},\qquad x_2=\frac{E-|{\bf
P}|\cos\psi}{\sqrt{s}} \; .
\end{eqnarray}
Relations (\ref{Epsi}) and (\ref{x1x2cos}) show that the interchange
of proton momenta $p_1\leftrightarrow p_2$ is equivalent to the
interchange of the angle $\psi\leftrightarrow\psi\pm\pi$, i.e.
  $\sin\psi\leftrightarrow-\sin\psi$ and
$\cos\psi\leftrightarrow-\cos\psi$  simultaneously. The last
permutation also provides the interchange of the longitudinal
components of gluons momenta $x_1\leftrightarrow x_2$.

Conservation laws provide us with the following relations between
components of gluon transverse momenta and covariant scalar products
\begin{eqnarray}\nonumber
&&Q_{1,t}^{x}=-\frac{q_{1,t}^2+(q_{1,t}q_{2,t})}{|{\bf
P}|\sin\psi},\quad
Q_{2,t}^{x}=-\frac{q_{2,t}^2+(q_{1,t}q_{2,t})}{|{\bf
P}|\sin\psi},\quad
Q_t^{y}=\frac{\sqrt{q_{1,t}^2q_{2,t}^2-(q_{1,t}q_{2,t})^2}}{|{\bf
P}_t|}\,
\mathrm{sign}(Q_t^y),\label{rel_comp} \\
&&P_{t}^2=-|{\bf P}_t|^2=-|{\bf
P}|^2\sin^2\psi=q_{1,t}^2+q_{2,t}^2+2(q_{1,t}q_{2,t}),\quad
q_{1/2,t}^2=-|{\bf q}_{1/2,t}|^2, \nonumber
\end{eqnarray}
where $|{\bf P}_t|=|{\bf P}||\sin\psi|$ is the meson transverse
momentum with respect to the $z$-axis. The appearance of the factor
$\mathrm{sign}(Q_t^y)$ guarantees the applicability of
(\ref{rel_comp}) for positive and negative $Q_t^y$. Note that under
permutations $q_{1,t}\leftrightarrow q_{2,t}$ implied by Bose
statistics the components interchange as $Q_{1,t}^{x}\leftrightarrow
Q_{2,t}^{x}$ and $Q_{t}^{y}\leftrightarrow -Q_{t}^{y}$. In our
notations the quantity $\sin\psi$ plays a role of the
noncollinearity of meson in considered coordinates. A
straightforward calculation leads to the following vertex function
in these coordinates
\begin{eqnarray}\label{chi2-fin}
&&V^{c_1c_2}_{J=2,\,\lambda}=2ig^2\delta^{c_1c_2}\sqrt{\frac{1}{3M\pi
N_c}}\frac{{\cal R}'(0)}{M|{\bf
P}_t|^2(M^2-q_{1,t}^2-q_{2,t}^2)^2}\times\\
&&\biggl[6M^2\,i|\lambda|(q_{1,t}^2-q_{2,t}^2)\,\mathrm{sign}(Q_t^y)\Big\{|[{\bf
q}_{1,t}\times{\bf q}_{2,t}]\times{\bf
n}_1|\,(1-|\lambda|)\,\mathrm{sign}(\sin\psi)\,\mathrm{sign}(\cos\psi)+\nonumber\\
&&2\,|[{\bf q}_{1,t}\times{\bf q}_{2,t}]\times{\bf
n}_3|\,(2-|\lambda|)\Big\}-\big[2q_{1,t}^2q_{2,t}^2+(q_{1,t}^2+q_{2,t}^2)(q_{1,t}q_{2,t})\big]
\Big\{3M^2(\cos^2\psi+1)\lambda(1-|\lambda|)+\nonumber\\
&&6ME\sin(2\psi)\,\lambda(2-|\lambda|)\,\mathrm{sign}(\sin\psi)\,\mathrm{sign}(\cos\psi)+
\sqrt{6}\,(M^2+2E^2)\sin^2\psi\,(1-|\lambda|)(2-|\lambda|)\Big\}\biggr]
\; ,
\nonumber
\end{eqnarray}
where
\begin{eqnarray*}
&&|[{\bf q}_{1,t}\times{\bf q}_{2,t}]\times{\bf
n}_1|=\sqrt{q_{1,t}^2q_{2,t}^2-(q_{1,t}q_{2,t})^2}\,|\cos\psi|,\\
&&|[{\bf q}_{1,t}\times{\bf q}_{2,t}]\times{\bf
n}_3|=\frac{E}{M}\sqrt{q_{1,t}^2q_{2,t}^2-(q_{1,t}q_{2,t})^2}\,|\sin\psi|.
\end{eqnarray*}
The amplitude (\ref{chi2-fin}) explicitly obeys the Bose symmetry
under the interchange of gluon momenta and polarizations due to
resulting simultaneous permutations
$\cos\psi\leftrightarrow-\cos\psi$,
$\sin\psi\leftrightarrow-\sin\psi$ and $Q_{t}^{y}\leftrightarrow
-Q_{t}^{y}$.

It follows from the conservation laws that
\begin{eqnarray}\nonumber
q_{1t}+p'_{1t}=-q_{0t},\qquad q_{2t}+p'_{2t}=q_{0t},\qquad
P_t=-(p'_{1t}+p'_{2t})
\end{eqnarray}
Let us consider first the limit of the ``coherent'' scattering of
protons $p'_{1t}=p'_{2t}\equiv p_t$, so
\begin{eqnarray}\label{coher-mom}
q_{1t}=-(p_t+q_{0t}),\qquad q_{2t}=-(p_t-q_{0t}),\qquad
P_t=-2p_{t},\qquad p_t^y=0.
\end{eqnarray}
The production vertex (\ref{chi2-fin}) in this limit has the form
\begin{eqnarray}\nonumber
&&V^{c_1c_2}_{J=2,\,\lambda}(q_{0t}^x,q_{0t}^y,p_t)=2ig^2\delta^{c_1c_2}\sqrt{\frac{1}{3M\pi
N_c}}\frac{{\cal R}'(0)}{M(M^2-2(p_t^2+q_{0t}^2))^2}\times\\
&&\biggl[12M^2\,i|\lambda|\,q_{0t}^xq_{0t}^y\Big\{(1-|\lambda|)
\,\cos\psi+\frac{2E}{M}\,(2-|\lambda|)\,\sin\psi\Big\}+\label{chi2-fin-coher}\\
&&\big[p_t^2+(q_{0t}^x)^2-(q_{0t}^y)^2\big]
\Big\{3M^2(\cos^2\psi+1)\lambda(1-|\lambda|)+\nonumber\\
&&6ME\sin(2\psi)\,\lambda(2-|\lambda|)\,\mathrm{sign}(\sin\psi)\,\mathrm{sign}(\cos\psi)+
\sqrt{6}\,(M^2+2E^2)\sin^2\psi\,(1-|\lambda|)(2-|\lambda|)\Big\}\biggr]
\nonumber
\end{eqnarray}
We see that in contrast to the axial-vector case considered in
Ref.~\cite{Pasechnik:2009bq}, the diffractive amplitude of
$\chi_c(2^+)$ production does not turn to zero in this ``coherent''
limit for $p_t\not=0$.

In the forward limit $p_{t}\to0$ (which is a particular case of
the ``coherent'' one) the amplitude turns to zero at any meson
rapidities $y$. Indeed, we have $P_t\to0$ and $\sin\psi\to\pm0$ and
the amplitude turns into
\begin{eqnarray}\label{chi2-forw}
V^{c_1c_2}_{J=2,\,\lambda}(q_{0t}^x,q_{0t}^y,p_t\to0)&=&g^2\delta^{c_1c_2}\sqrt{\frac{1}{3M\pi
N_c}}\frac{12M\,{\cal R}'(0)}{(M^2-2q_{0t}^2)^2}\times\\
&&(1-|\lambda|)\big\{i\lambda\big[(q_{0t}^x)^2-(q_{0t}^y)^2\big]-2\,|\lambda|\,q_{0t}^xq_{0t}^y\,
\mathrm{sign}(\cos\psi)|_{\psi\to0,\pi}\big\} \nonumber \, .
\end{eqnarray}
The imaginary part of this vertex function turns out to be
antisymmetric w.r.t. interchanging $q_{0t}^x\leftrightarrow
q_{0t}^y$, whereas its real part is antisymmetric w.r.t. changing
the sign of $q_{0t}^x$ or $q_{0t}^y$ component, i.e.
\begin{eqnarray*}
&&\Im V^{c_1c_2}_{J=2,\,\lambda}(q_{0t}^x,q_{0t}^y,p_t\to0)= -\Im
V^{c_1c_2}_{J=2,\,\lambda}(q_{0t}^y,q_{0t}^x,p_t\to0)\,\\
&&\Re V^{c_1c_2}_{J=2,\,\lambda}(q_{0t}^x,q_{0t}^y,p_t\to0)=-\Re
V^{c_1c_2}_{J=2,\,\lambda}(-q_{0t}^x,q_{0t}^y,p_t\to0)=-\Re
V^{c_1c_2}_{J=2,\,\lambda}(q_{0t}^x,-q_{0t}^y,p_t\to0)\,.
\end{eqnarray*}
Since in this case $q_{1t}=-q_{0t},\,q_{2t}=q_{0t}$ in the forward
limit, then the double integral in the diffractive amplitude has an
antisymmetric integrand and turns to zero in the symmetric limit
\begin{eqnarray}\nonumber
{\cal M}_{p_t\to0}\sim F_1(t_1)F_1(t_2)\int
dq_{0t}^xdq_{0t}^y\frac{V_{J=2}(q_{0t}^x,q_{0t}^y,p_t\to0)\cdot
f(x_1,q_{0,t}^2,q_{0,t}^2)f(x_2,q_{0,t}^2,q_{0,t}^2)}{q_{0t}^6}=0.\\
\label{forw}
\end{eqnarray}
This explicitly confirms the observation made in
Refs.~\cite{KKMR-spin,Yuan01}\footnote{We are grateful to
V.~A.~Khoze for helpful discussions of this problem.}.

Very recently, when our paper was almost complete, a paper by L.
Harland-Lang, V. Khoze, M. Ryskin and W. Stirling (HKRS)
\cite{HarlandLang:2009qe} appeared where the hard subprocess
amplitudes $gg\to \chi_c(J^+)$ (based on formalism by Kuhn et al for
$\gamma^*\gamma^*\to\chi_{c}(J^{+})$ \cite{Kuhn-g}) including the
gluon virtualities were listed for different spins including the
tensor $\chi_c(2^+)$:
\begin{eqnarray}
V_{J=0}^{\mathrm{HKRS}}&=&\sqrt{\frac16}\frac{c}{M}\big[3M^2(q_{1,t}q_{2,t})-
(q_{1,t}q_{2,t})(q_{1,t}^2+q_{2,t}^2)-2q_{1,t}^2q_{2,t}^2\big],\label{HKRS-J0}\\
V_{J=1,\lambda}^{\mathrm{HKRS}}&=&-\frac{2ic}{s}p_{1,\nu}p_{2,\alpha}\varepsilon^{\mu\nu\alpha\beta}\epsilon_{\beta}
\big[(q_{2,t})_{\mu}q_{1,t}^2-(q_{1,t})_{\mu}q_{2,t}^2\big],\label{HKRS-J1}\\
V_{J=2,\lambda}^{\mathrm{HKRS}}&=&\frac{\sqrt{2}cM}{s}\,\epsilon^{\mu\alpha}
\big[s(q_{1,t})_{\mu}(q_{2,t})_{\alpha}+2(q_{1,t}q_{2,t})p_{1,\mu}p_{2,\alpha}\big],
\label{HKRS-J2}
\end{eqnarray}
where the constant prefactor is
\[c=\frac{1}{2\sqrt{N_c}}\frac{4g^2}{(q_1q_2)^2}\sqrt{\frac{6}{4\pi M}}{\cal R}'(0)\,.\]

The first amplitude $V_{J=0}^{\mathrm{HKRS}}$ (\ref{HKRS-J0}) is the
same as the expression obtained in Ref.~\cite{Pasechnik:2007hm} (up
to a factor of 2 coming from different normalisations of the hard
part $n^+_{\mu}n^-_{\nu}V_{J,\,\mu\nu}$ in our case and
$(2/s)p_1^{\mu}p_2^{\nu}V_{J,\,\mu\nu}$ in
Ref.~\cite{HarlandLang:2009qe}), where the major role of the gluon
virtualities in the hard subprocess amplitude of quarkonia
production was claimed to be crucial. In particular, it was shown
that an account of the gluon virtualities reduces the previous KMRS
result in Ref.~\cite{Khoze:2004yb} for on-mass-shell gluons
$V^{\mathrm{KMRS}}_0\sim (q_{1,t}q_{2,t})$ by a factor of 2 -- 3.

The second amplitude, $V_{J=1,\lambda}^{\mathrm{HKRS}}$, looks
different from our previous result, obtained in
Ref.~\cite{Pasechnik:2009bq}. However, one can directly check that
the difference between the amplitudes (\ref{HKRS-J1}) and (2.12) in
Ref.~\cite{Pasechnik:2009bq} turns to zero when fixing the
coordinates in the c.m.s. frame of reference as in
Eq.~(\ref{protons}) (see also Fig.~\ref{fig:cms}) and the meson
polarisation vector $\epsilon^{\beta}$ with the basis as in
Eq.~(\ref{basis}). Due to the covariant structure of these
amplitudes, the last observation means that they are the same in any
frame of reference. The calculations proving this equality are
rather involved, and we do not show them explicitly here.

Very similar situation holds for $\chi_c(2^+)$ production
amplitudes. Namely, the amplitudes (\ref{HKRS-J2}) and
(\ref{Vgen-chi2}) turned out to be the same under fixing the
coordinates as in the previous section. Therefore, under the
kinematical relations our results for the hard subprocess amplitudes
are in complete agreement with the corresponding HKRS results. Let
us now turn to the discussion of numerical results.

\section{Numerical results}

Results for the differential cross sections $d\sigma/dy(y=0)$ of
the diffractive $\chi_c(0^+,1^+,2^+)$ meson production at the Tevatron
energy $W = 1960$ GeV for different UGDFs are shown in
Table~\ref{table:dy0}. In the last column we show the results for
the expected signal in the $J/\psi+\gamma$ channel summed over all
$\chi_c$ spin states (and all polarisation states of
$\chi_c(1^+,2^+)$ mesons)
\begin{eqnarray}
\frac{d\sigma_{obs}}{dy}\Big|_{y=0} \approx \sum_{J=0,1,2}
K^{(J)}_{\mathrm{NLO}}\langle S_{\mathrm{eff}}^2\rangle_J
\mathrm{BR}(\chi_c(J^+) \to J/\Psi+\gamma)
\frac{d\sigma^{bare}_{\chi_c(J^+)}}{dy}\Big|_{y=0}, \label{sig}
\end{eqnarray}
which can be compared with the corresponding value
measured by the CDF Collaboration \cite{Aaltonen:2009kg}:
$d\sigma^{exp}/dy|_{y=0}(pp\to
pp(J/\psi+\gamma))\simeq(0.97\pm0.26)$ nb.

In Refs.~\cite{Khoze:2004yb,HarlandLang:2009qe} is was assumed that
the NLO corrections factor $K_{\mathrm{NLO}}$ in the $g^*g^*\to
\chi$ vertex is the same as in the $\chi\to gg$ width implying that
$|V_J|^2\sim\Gamma(\chi\to gg)$. In general, such corrections depend
on spin of $q\bar{q}$ resonance. So, the diffractive cross section
for each $\chi_c(J^{+})$ has to be multiplied by not necessarily the
same factor $K^{(J)}_{\mathrm{NLO}}$, as shown in
Eq.~(\ref{sig})\footnote{See Ref.~\cite{HarlandLang:2009qe} for
discussion of extra uncertainties coming from NNLO and higher order
corrections.}. This can be done, however, only for $0^{+}$ and
$2^{+}$ states, and the corresponding NLO QCD radiative corrections
are well-known \cite{qcdcorr}:
\begin{eqnarray}\label{knlo}
K^{(0)}_{\mathrm{NLO}}=1+8.77\frac{\alpha_s(M_{\chi})}{\pi}\simeq
1.68,\qquad
K^{(2)}_{\mathrm{NLO}}=1-4.827\frac{\alpha_s(M_{\chi})}{\pi}\simeq
0.63\,.
\end{eqnarray}
Due to the Landau-Yang theorem the decay of the axial vector
charmonium $1^{++}$ to on-shell gluons is forbidden, and there are
no reliable calculations of the NLO QCD corrections to its coupling with
off-shell gluons. In the following we take naively
$K^{(1)}_{\mathrm{NLO}}=1$. This leads to an additional
uncertainty of the model predictions.

As has been claimed in Refs.~\cite{Martin:2009ku,HarlandLang:2009qe}
the absorptive corrections are quite sensitive to the meson
spin-parity. This was studied before in the context of scalar and
pseudoscalar Higgs production in Ref.~\cite{KKMR}. We adopt here the
following effective gap survival factors, calculated in
Ref.~\cite{HarlandLang:2009qe} for different spins including eikonal
and so-called enhanced contributions:
\begin{eqnarray} \langle
 S_{\mathrm{eff}}^2(\chi_c(0^+))\rangle\simeq0.033,\qquad \langle
 S_{\mathrm{eff}}^2(\chi_c(1^+))\rangle\simeq0.050,\qquad \langle
 S_{\mathrm{eff}}^2(\chi_c(2^+))\rangle\simeq0.073\,.
 \label{S2eff}
\end{eqnarray}

The contribution of the scalar $\chi_c(0^+)$ CEP, which was
initially assumed to be the dominant one \cite{Khoze:2004yb}, is
reduced by a very small branching ratio of its observable radiative
decay \cite{Pasechnik:2009bq,HarlandLang:2009qe}. In turn, the
strong suppression of the $\chi_c(1^+)$ central production in both
the on-mass-shell limit of fusing gluons (due to Landau-Yang theorem
\cite{LY_theorem}) and the forward scattering limit of outgoing
protons (due to the so-called $J_z=0$ selection rule
\cite{Khoze:2004yb,Khoze:2000jm}) may be partially compensated by
its much higher branching ratio to the observed $J/\psi+\gamma$
final state \cite{Pasechnik:2009bq}. Analogously to the axial-vector
case, the suppression of the tensor $\chi_c(2^+)$ CEP is likely to
be eliminated by its large decay branching ratio
\cite{HarlandLang:2009qe}, and the resulting value of the radiative
decay signal is under our special interest.
\begin{table}[!h]
\caption{\label{table:dy0} \small\sf Differential cross section
$d\sigma_{\chi_c}/dy(y=0)$ (in nb) of the exclusive diffractive
production of $\chi_c(0^+,1^+,2^+)$ mesons and their partial and
total signal in radiative $J/\psi+\gamma$ decay channel
$d\sigma_{J/\psi\gamma}/dy(y=0)$ at Tevatron for different UGDFs,
cuts on the transverse momentum of the gluons in the loop
($q_{0,t}$) and different values of the auxiliary parameter $\xi$
controlling the characteristic $x'$ values in the symmetric skewed
UGDFs prescription (\ref{skewed-UGDFs}) (denoted as ``sqrt''). NLO
skewedness factor $R^{NLO}_g=1.3$ for the KMR asymmetric
prescription (\ref{asym-off}) (denoted as ``$R_{g}$''), NLO
correction factors (\ref{knlo}) and absorptive correction factors
(\ref{S2eff}) are included. Contributions from all polarisations are
incorporated.}
 {\small
\begin{center}
\begin{tabular}{|c|c||c|c||c|c||c|c||c|c|c||}
\hline
 skewed UGDF & &\multicolumn{2}{l||}{$\quad\;\;\chi_c(0^+)\quad\quad$}&
  \multicolumn{2}{l||}{$\quad\;\;\chi_c(1^+)\quad\quad$}&
  \multicolumn{2}{l||}{$\quad\;\;\chi_c(2^+)\quad\quad$}& \multicolumn{2}{l|}{$\quad\quad$ratio} & signal \\
\cline{3-10}
 prescription & $\;\;\xi\;\;$ &$\;\,\frac{d\sigma_{\chi_c}}{dy}\;\,$&$\frac{d\sigma_{J/\psi\gamma}}{dy}$&
 $\;\,\frac{d\sigma_{\chi_c}}{dy}\;\,$&$\frac{d\sigma_{J/\psi\gamma}}{dy}$&
 $\;\,\frac{d\sigma_{\chi_c}}{dy}\;\,$&$\frac{d\sigma_{J/\psi\gamma}}{dy}$&
 $1^+/0^+$
 & $2^+/0^+$ &
 $\frac{d\sigma_{obs}}{dy}$ \\
\hline\hline
 GBW \cite{GBW}, ``sqrt''
                                         & 1.0 & 13.2 & 0.15 & 0.01  & 0.003 & 1.96  & 0.38 & 0.02 & 2.5 & 0.5  \\ \cline{2-11}
                                         & 0.3 & 12.8 & 0.15 & 0.04  & 0.01 & 1.39  & 0.27 & 0.07 & 1.8 & 0.4  \\
\hline\hline
 lin KS \cite{KS05}, ``$R_g$''           & --- & 32.6 & 0.37 & 0.20  & 0.07 & 0.53  & 0.10 & 0.2 & 0.3 & 0.5  \\ \hline
 lin KS \cite{KS05}, ``sqrt''            & 1.0 & 17.2 & 0.19 & 0.12  & 0.04 & 0.61  & 0.12 & 0.2 & 0.6 & 0.4  \\ \hline\hline
 nlin KS \cite{KS05}, ``sqrt''           & 1.0 & 12.6 & 0.14 & 0.07  & 0.02 & 0.39  & 0.08 & 0.1 & 0.6 & 0.3  \\ \cline{2-11}
                                         & 0.3 & 20.6 & 0.23 & 0.10  & 0.03 & 0.57  & 0.11 & 0.1 & 0.5 & 0.4  \\ \cline{2-11}
                                         & $\bm{0.05}$ & $\bm{36.0}$ & $\bm{0.41}$ & $\bm{0.13}$  & $\bm{0.04}$ & $\bm{0.84}$
                                         & $\bm{0.16}$ & $\bm{0.1}$ & $\bm{0.4}$ & $\bm{0.6}$  \\
 \hline\hline
 KMR \cite{MR}, GRV94HO,                 &     &      &      &      &      &      &      &     &     &      \\
 $(q^{cut}_{0,t})^2=0.72\,\GeV^2$              & ---  & 13.5 & 0.16 & 0.06 & 0.02 & 0.23 & 0.05 & 0.1 & 0.3 & 0.2  \\ \hline
 KMR \cite{MR}, GRV94HO,                 &     &      &      &      &      &      &      &     &     &      \\
 $(q^{cut}_{0,t})^2=0.36\,\GeV^2$              & --- & $\bm{37.9}$ & $\bm{0.43}$ & $\bm{0.14}$ & $\bm{0.05}$
                                               & $\bm{0.75}$ & $\bm{0.15}$ & $\bm{0.1}$  & $\bm{0.3}$ & $\bm{0.6}$  \\
 \hline\hline
 HKRS result \cite{HarlandLang:2009qe}   &     &      &      &      &      &      &      &     &     &      \\
 $(q^{cut}_{0,t})^2=0.72\,\GeV^2$              & --- & 27.1 & 0.31 & 0.72 & 0.25 & 0.95 & 0.19 & 0.8 & 0.6 & 0.7  \\
 \hline
\end{tabular}
\end{center}}
\end{table}

As it was discussed in Ref.~\cite{Pasechnik:2007hm}, the dominant
contribution to the diffractive CEP of $\chi_c(0^+)$ comes from
nonperturbative values of the gluon transverse momenta
$q_t<1\,\GeV$. In order to estimate the role of small $q_t$ in the
central production of $\chi_c(1^+,2^+)$ and related theoretical
uncertainties we use different UGDFs known from the literature (for
details see Refs.~\cite{LS06,Pasechnik:2007hm}). Among them there
are perturbatively modeled KMR UGDF \cite{KMR,Kimber:2001sc,MR},
which include the Sudakov form factor, as well as GBW \cite{GBW} and
linear/nonlinear Kutak-Sta\'sto (KS) \cite{KS05} UGDF models which
by construction can be used for any values of the gluon transverse
momenta.

In the last row of Table~\ref{table:dy0} we show the HKRS results
for partial cross sections extracted from their original paper
\cite{HarlandLang:2009qe}. These cross sections were calculated by
the HKRS at some small energy scale and subsequently extrapolated up
to the Tevatron energy assuming a Regge type energy dependence.

By direct calculation at the Tevatron energy with the same KMR
UGDFs, but without imposing any arguments beyond the QCD framework
(like Regge scaling, for example), we get the observable
$J/\psi\gamma$ cross section
$d\sigma_{{J/\psi\gamma}}/dy(y=0)\simeq0.6$ nb, which is close to
HKRS result $\sim0.7$ nb, but somewhat lower than the present CDF
result $d\sigma^{exp}/dy|_{y=0}(pp\to
pp(J/\psi+\gamma))\simeq(0.97\pm0.26)$ nb. However, this demanded to
incorporate physics below HKRS cut-off on gluon transverse momentum
in the loop integral (\ref{ampl}) $(q^{cut}_{0,t})^2=0.72\,\GeV$
underlining the importance of nonperturbative contributions of small
$q_{0,t}$ in the QCD mechanism under consideration. Relations
between different $\chi_c$'s obtained in
Ref.~\cite{HarlandLang:2009qe} are not reproduced as well. Our
result is highlighted in bold in Table~\ref{table:dy0}.

The reason of such a discrepancy in both the normalization and
relative contributions of different spins to the observable rate is
the on-shell approximation $M_X^2\gg q_{1/2,t}^2$ in the hard
$g^*g^*\to\chi_c(J^+)$ adopted by HKRS\footnote{We are most thankful
to L. Harland-Lang for the helpful discussions and correspondence on
the topic. The exchange of the Fortran codes between us helped a lot
in cross-checks of our calculations and understanding the various
sources of discrepancies.}. As was firstly noticed in
Ref.~\cite{Pasechnik:2007hm}, the gluon off-shellness plays an
important role in the central exclusive $\chi_c(0^+)$ leading to a
strong reduction of the cross section depending on UGDF model. Now,
we observe that the same effect significantly affects the ratios
between different spin contributions.

Note that at zeroth meson rapidity $y=0$ a significant part of the
cross section comes from lower polarisation states in the
center-of-mass frame $\lambda=0$ ($\chi_c(1^+)$) and
$\lambda=0,\,\pm1$ ($\chi_c(2^+)$). In the total (integrated over
$y$) cross section the maximal helicity contributions, however,
strongly dominate. We leave a more detailed investigation of the
polarisation effects for a separate publication.
\begin{figure}[h!]
 \centerline{\epsfig{file=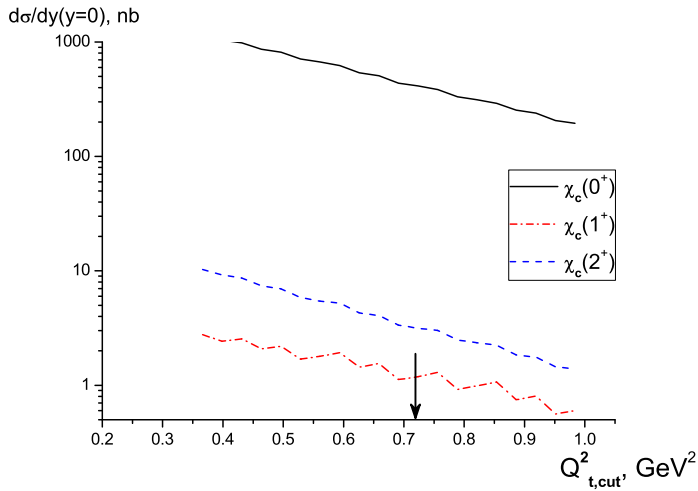,height=7cm,width=10cm}}
 \caption{\small \em Dependence of the differential cross section
 $d\sigma_{\chi_c}/dy(y=0)$ of $\chi_c(0^+,1^+,2^+)$ CEP
 on the infrared cut-off on small effective
 gluon transverse momentum $Q^{cut}_t$ for the KMR UGDF with GRV94HO
 ($R_g=1.3$). Absorption effects
 are not included here.
 Arrow points to the HKRS cut-off $0.72\,\GeV^2$ \cite{HarlandLang:2009qe}.}
 \label{fig:cutdep}  \end{figure}

Relative contributions of $\chi_c(0^+,1^+,2^+)$ CEP to observable
signal ($J/\Psi + \gamma$) require an additional discussion. Last
PDG updated set of branching ratios for charmonia radiative decays
is \cite{PDG}:
\begin{eqnarray*}
&&\mathrm{BR}(\chi_c(0^+)\to J/\psi+\gamma)=0.0114,\\
&&\mathrm{BR}(\chi_c(1^+)\to J/\psi+\gamma)=0.341,\\
&&\mathrm{BR}(\chi_c(2^+)\to J/\psi+\gamma)=0.194.
\end{eqnarray*}

Furthermore, as one can see in Table~\ref{table:dy0}, despite of
larger branching ratio in the axial-vector case the observable
signal from the $\chi_c(1^+)$ CEP occurs to be smaller than that
from $\chi_c(2^+)$ for UGDFs enhanced at sufficiently small
nonperturbative $q_t$ (in particular, for the Kutak-Sta\'sto (KS)
and GBW UGDFs) due to an additional suppression of the $g^*g^*\to
\chi_c(1^+)$ subprocess vertex at small $q_{1/2,t}$. For the GBW
UGDF the $\chi_c(1^+)$ contribution is strongly suppressed whereas
the $\chi_c(2^+)$ contribution turned out to be larger than the
$\chi_c(0^+)$ one. All the UGDF models under considerations lead to
somewhat underestimated observable signal at Tevatron $\lesssim0.6$
nb, however, it can be still reliable within relatively large
theoretical uncertainties of the gap survival factors and QCD
mechanism under discussion. In the case of the KS model,
contributions from $\chi_c(1^+)$ state are found to be stronger
suppressed than in KMR model. Measurements of separate spin
contributions thus would help to distinguish between different UGDF
models.
%
\begin{table}[!h]
\caption{\label{table:Edep} \small\sf  Integrated over full phase
space (bare) cross sections (in nb) for the central exclusive
$\chi_c(0^+,1^+,2^+)$ production at RHIC, Tevatron and LHC energies.
Infrared cut-off for the KMR UGDF (in ``$R_g$''-prescription) is
taken to be $(Q^{cut}_t)^2=0.72\,\GeV^2$. We take $R_g$ to be equal
1.3 at all three energies. Absorption effects are not included
here.}
\begin{center}
\begin{tabular}{|c||c||c|c|c|}
\hline
$\chi_c$ & UGDF &$\quad$RHIC$\quad$&$\quad$Tevatron$\quad$& $\quad$LHC$\quad$\\
\hline
 $\chi_c(0^+)$ & nlin. KS, $\xi=0.3$  & 108    & 3569  & 23260  \\
\cline{2-5}
               & KMR, GRV94HO         & 43     & 2270  & 30720  \\
\hline \hline
 $\chi_c(1^+)$ & nlin. KS, $\xi=0.3$  & 0.4    & 12   & 65     \\
\cline{2-5}
               & KMR, GRV94HO         & 0.2    & 7    & 87    \\
\hline \hline
 $\chi_c(2^+)$ & nlin. KS, $\xi=0.3$  & 2      & 44   & 209  \\
\cline{2-5}
               & KMR, GRV94HO         & 0.4    & 18   & 195  \\
\hline
\end{tabular}
\end{center}
\end{table}

In the case of the KMR UGDF, we observe quite substantial dependence
of the predicted observable signal w.r.t. variations of the infrared
cut-off on small transverse momenta of the gluons in the most
internal loop (see Fig.~\ref{fig:cutdep}). From
Table~\ref{table:dy0} we see that the shift of $Q^{cut}_t$ from the
value $0.72\,\GeV^2$ used in Ref.~\cite{HarlandLang:2009qe} down to
the minimal perturbative scale of the integrated GRV94HO
distributions $0.36\,\GeV^2$ \cite{GRV} leads to increase of the
cross section by a factor of about 3, approaching the CDF data. For
comparison, decrease of the $Q_{cut}$ from $1\,\GeV^2$ down to
$0.36\,\GeV^2$ leads to increase of the cross section by a factor of
6. Since we can not estimate the nonperturbative contribution coming
from below $0.36\,\GeV^2$, this allows us to conclude that
perturbatively motivated KMR UGDF leads to infrared unstable result
in the case of relatively light charmonium CEP. It is clear that the
essential part of the QCD dynamics comes from the nonperturbative
region of transverse momenta below the HKRS cut-off $0.72\,\GeV^2$
\cite{HarlandLang:2009qe}. KS and GBW UGDFs allow to incorporate
some unknown physics even below the minimal GRV scale
$Q_0^2=0.36\,\GeV^2$, avoiding ambiguities in defining the effective
gluon momenta.

Applying the KMR's asymmetrical off-diagonal UGDF according to
Eq.~(\ref{asym-off}) (``$R_g$'' prescription) in the case of the GBW
models we get strongly overestimated observable signal at Tevatron,
which means that in this case it is crucial to take into account the
$x'$-dependence of off-diagonal UGDFs when going deeply into the
infrared region of small $q_t$'s. The $x'$-dependent ``sqrt''
prescription, introduced in Eq.~(\ref{skewed-UGDFs}), leads to
observable signal, which is much closer to the experimental data.

The ``sqrt'' prescription, introduced in Eq.~(\ref{skewed-UGDFs}),
provides an agreement with the data (within a factor of 2 in overall
theoretical uncertainty between different UGDFs) with the KS (with
rather small $\xi\sim 0.05$) and GBW models giving the cross section
$d\sigma_{obs}/dy(y=0)\simeq0.5 - 0.6$ (see Table~\ref{table:dy0}).
This practically means that the smaller $q_{0,t}$ comes into the
game, the smaller $x'$ w.r.t. $q_{0,t}^2/s$ is required to get the
data description, providing one more argument about importance of
nonperturbative effects in charmonia CEP. The relative contributions
of different charmonium states in the $J/\psi+\gamma$ channel
(including absorption effects) in the case of, e.g., KS model are
found to be:
\begin{equation}
 \left( \frac{d\sigma^{\chi_{c0}}_{J/\psi\gamma}}{dy} \right)_{KS}:
 \left( \frac{d\sigma^{\chi_{c1}}_{J/\psi\gamma}}{dy} \right)_{KS}:
 \left( \frac{d\sigma^{\chi_{c2}}_{J/\psi\gamma}}{dy} \right)_{KS} =
       1:0.1:0.4\,.
 \label{rel}
\end{equation}
They are not affected by smaller $x'$ or nonlinear effects in this
model. As the normalization point we took the contribution of the
$\chi_c(0^+)$ meson CEP as was done in
Ref.~\cite{HarlandLang:2009qe}.

\begin{figure}[!h]
\begin{minipage}{0.32\textwidth}
 \centerline{\includegraphics[width=1.3\textwidth]{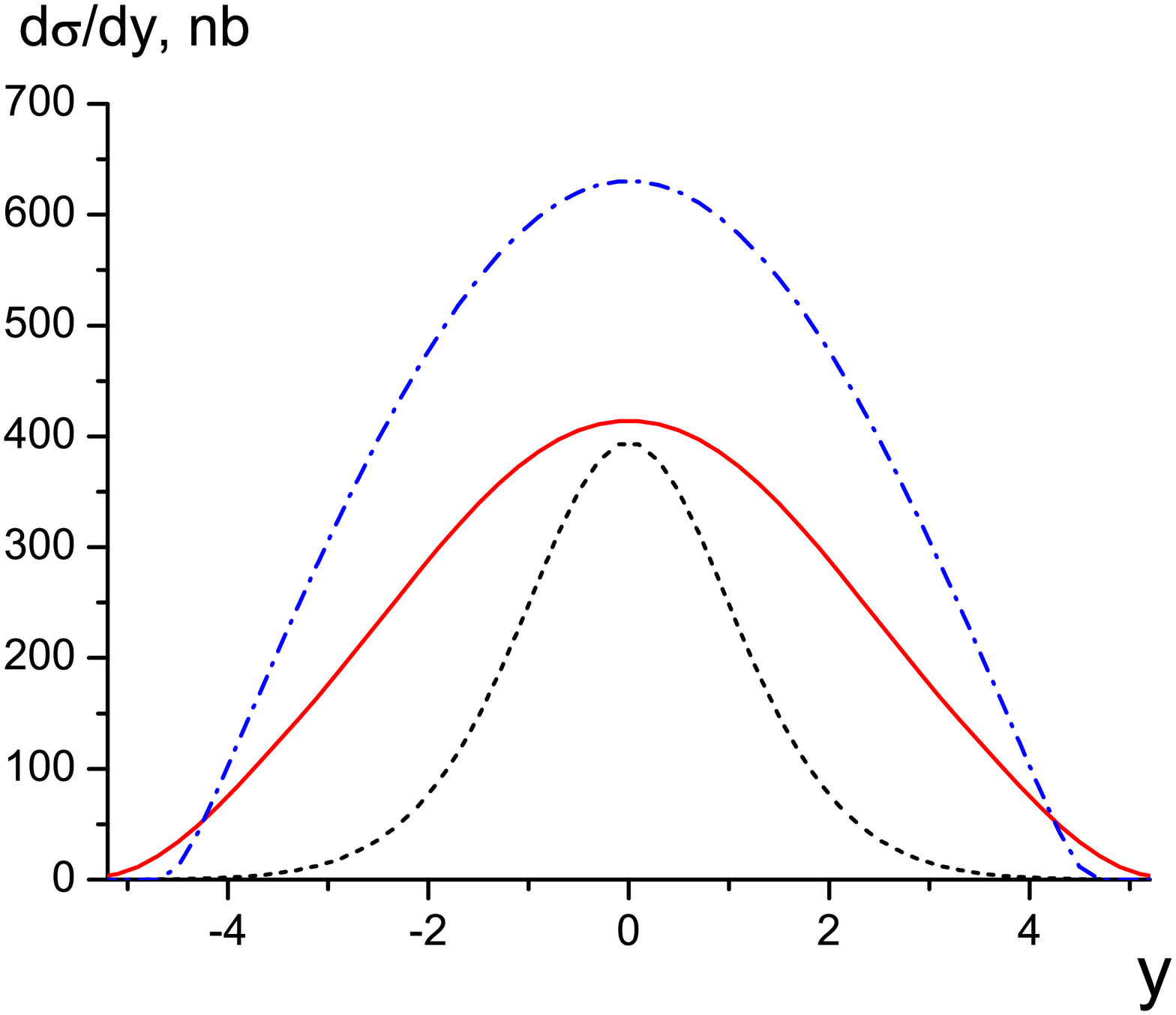}}
\end{minipage}
\begin{minipage}{0.32\textwidth}
 \centerline{\includegraphics[width=1.3\textwidth]{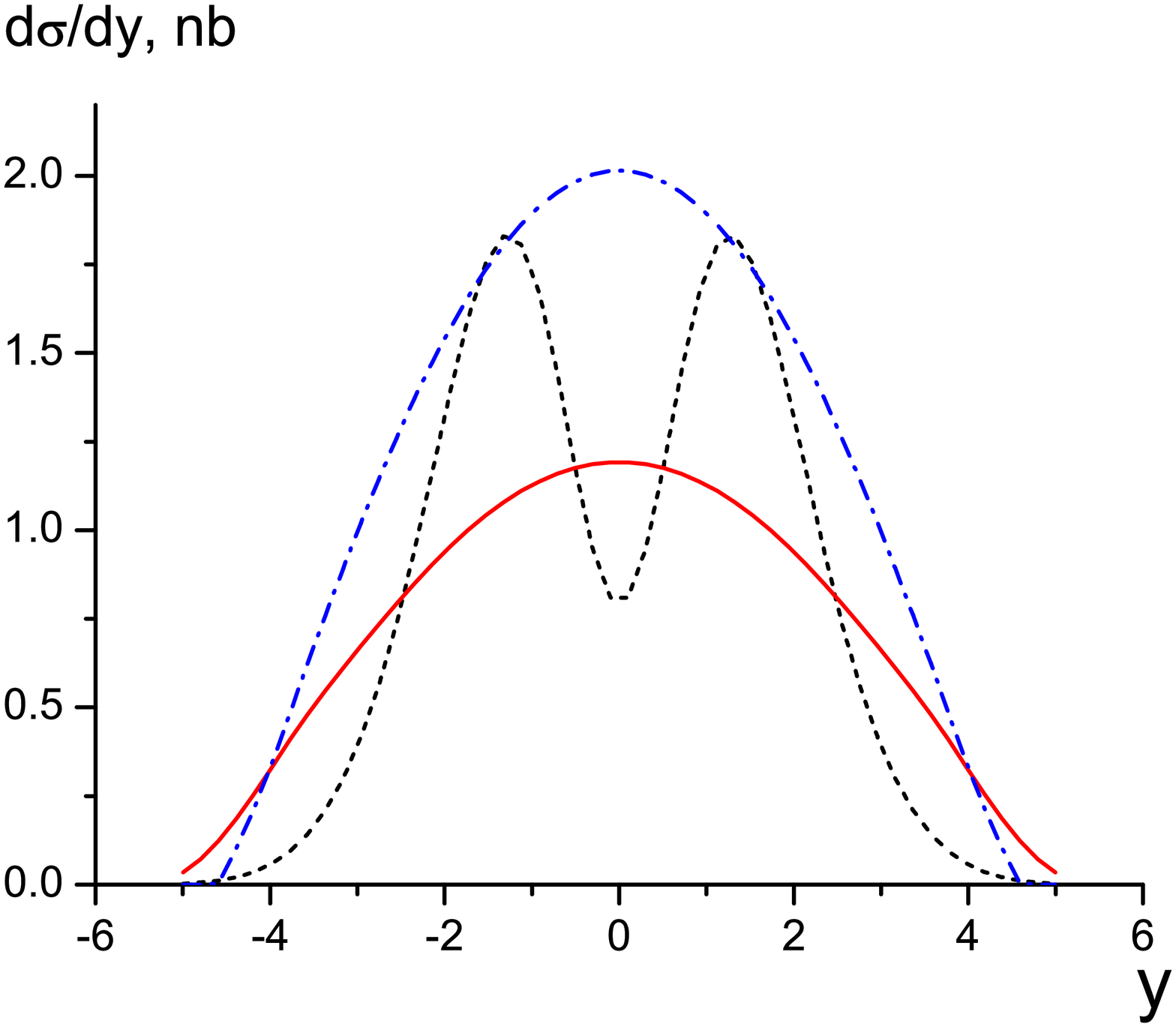}}
\end{minipage}
\begin{minipage}{0.32\textwidth}
 \centerline{\includegraphics[width=1.3\textwidth]{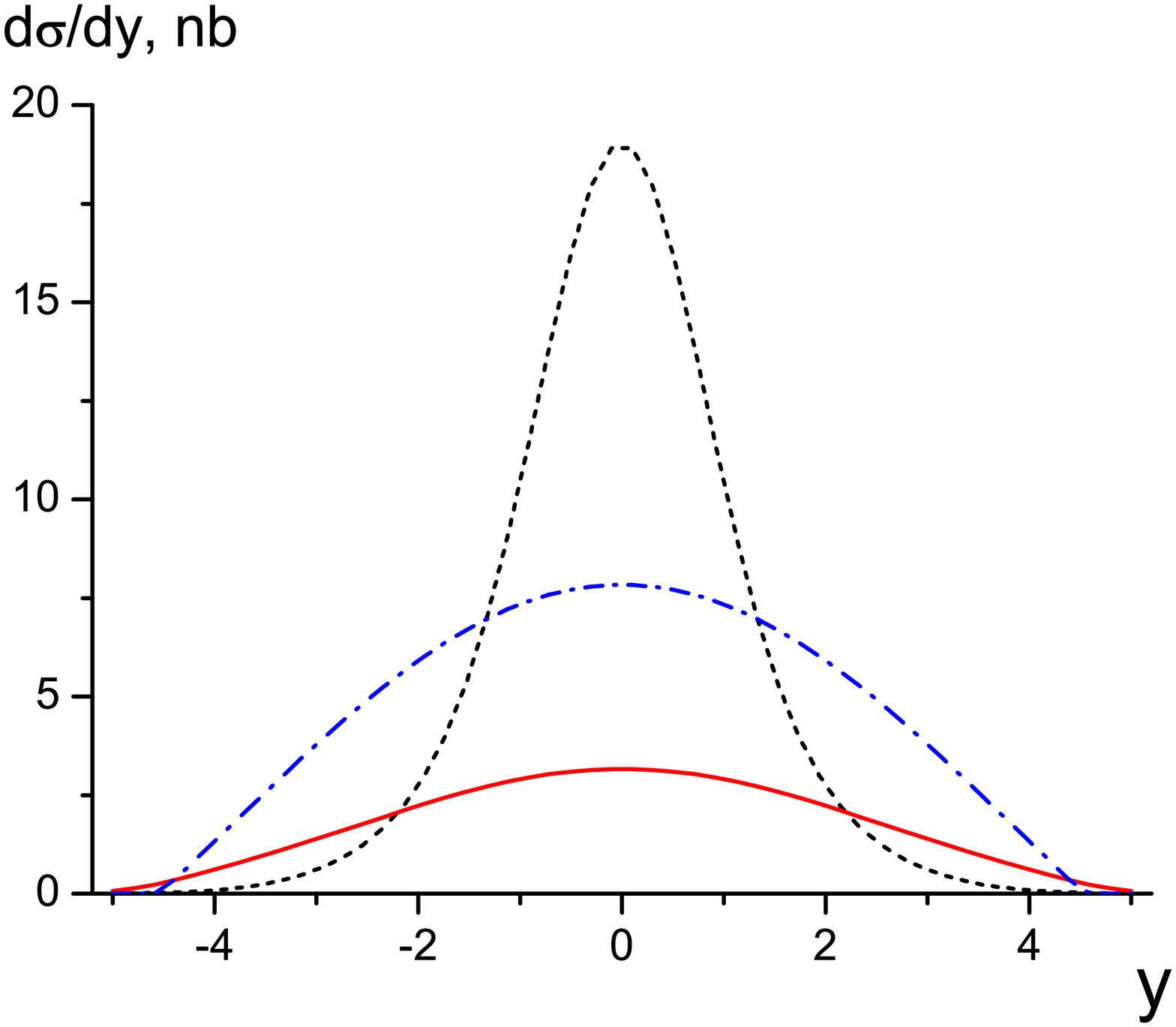}}
\end{minipage}
   \caption{\label{fig:chic012-dy}
   \small \em
Distributions $d\sigma_{\chi_c}/dy$ in rapidity of $\chi_c(0^+)$
(left panel), $\chi_c(1^+)$ (middle panel) and $\chi_c(2^+)$ (right
panel) mesons for different UGDFs at the Tevatron energy $\sqrt{s}$
= 1.96 TeV. The dash-dotted line corresponds to the KS UGDF
\cite{KS05} in the symmetrical ``sqrt''-prescription with $\xi=0.3$,
solid line -- KMR UGDF \cite{MR} with $R_g=1.3$,
$(Q^{cut}_t)^2=0.72\,\GeV^2$ and GRV94HO PDF \cite{GRV}, and
short-dashed line represents result with the GBW UGDF \cite{GBW}
($\xi=0.3$). Absorption effects are not included here.}
\end{figure}

In Table~\ref{table:dy0} we also presented results with the linear
Kutak-Sta\'sto model based on the unified BFKL-DGLAP framework and
the nonlinear one based on the Balitsky-Kovchegov equation
\cite{KS05}. It turned out that incorporation of the nonlinear
effects responsible for the gluon recombination in this model
reduces the $\chi_c(J^+)$ CEP cross sections by 30-50 \%. We see
that the nonlinear effects play a crucial role in diffractive
quarkonia production effectively decreasing the characteristic
values of $x'$ (controlled by $\xi$). However, reliable predictions
including the nonlinear effects require the exact knowledge of the
triple Pomeron vertex at NLLx accuracy, which is yet unknown.

It is also interesting to compare the diffractive production
of $\chi_c$ states at different energies. As an example, in
Table~\ref{table:Edep} we present the integrated (over full phase space)
cross sections of $\chi_c(0^+,1^+,2^+)$ production at RHIC, Tevatron
and LHC energies.
The results show similar energy behavior of the diffractive cross
section for different UGDFs as well as for different $\chi_c$ states.

\begin{figure}[!h]
\begin{minipage}{0.32\textwidth}
 \centerline{\includegraphics[width=1.3\textwidth]{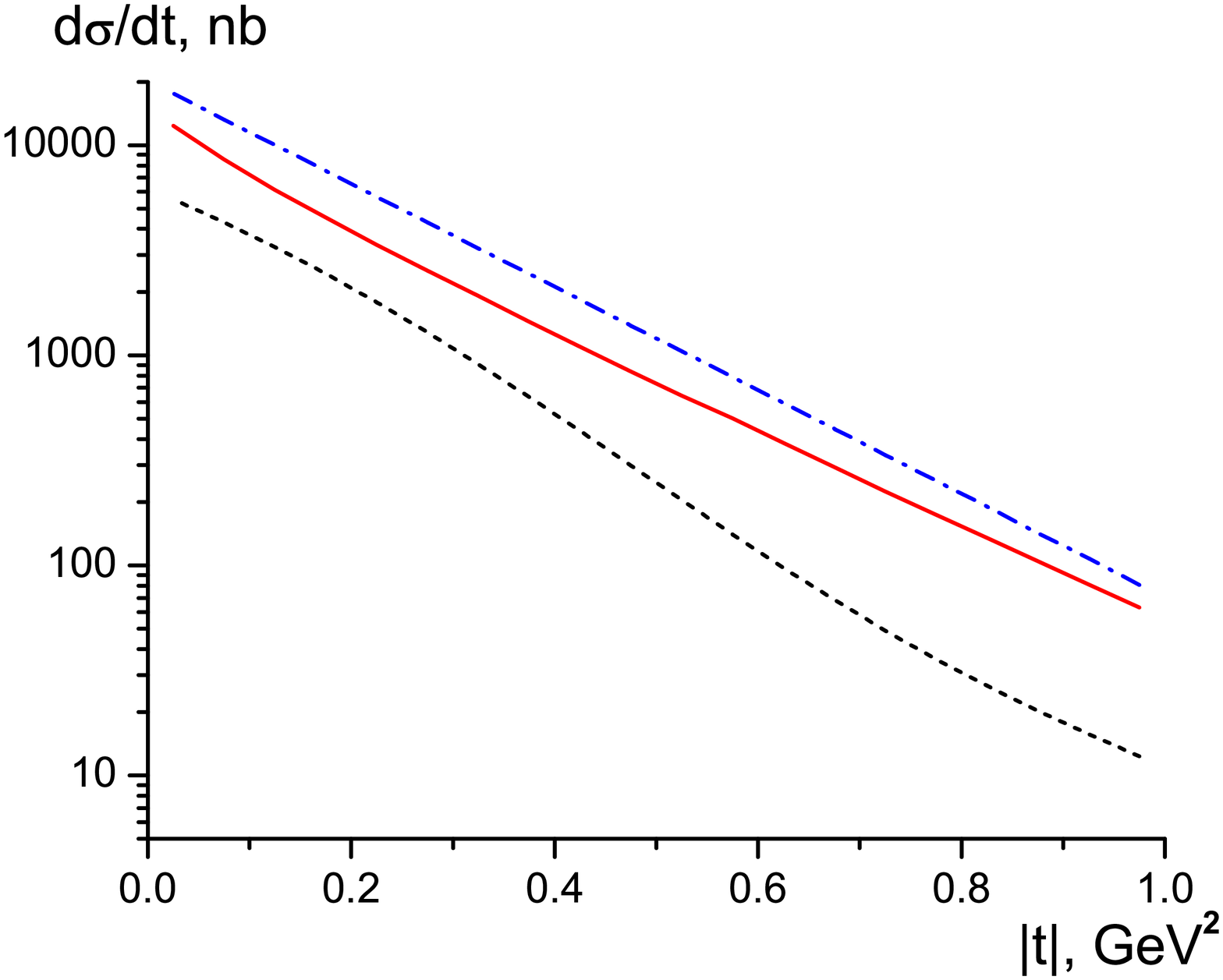}}
\end{minipage}
\begin{minipage}{0.32\textwidth}
 \centerline{\includegraphics[width=1.3\textwidth]{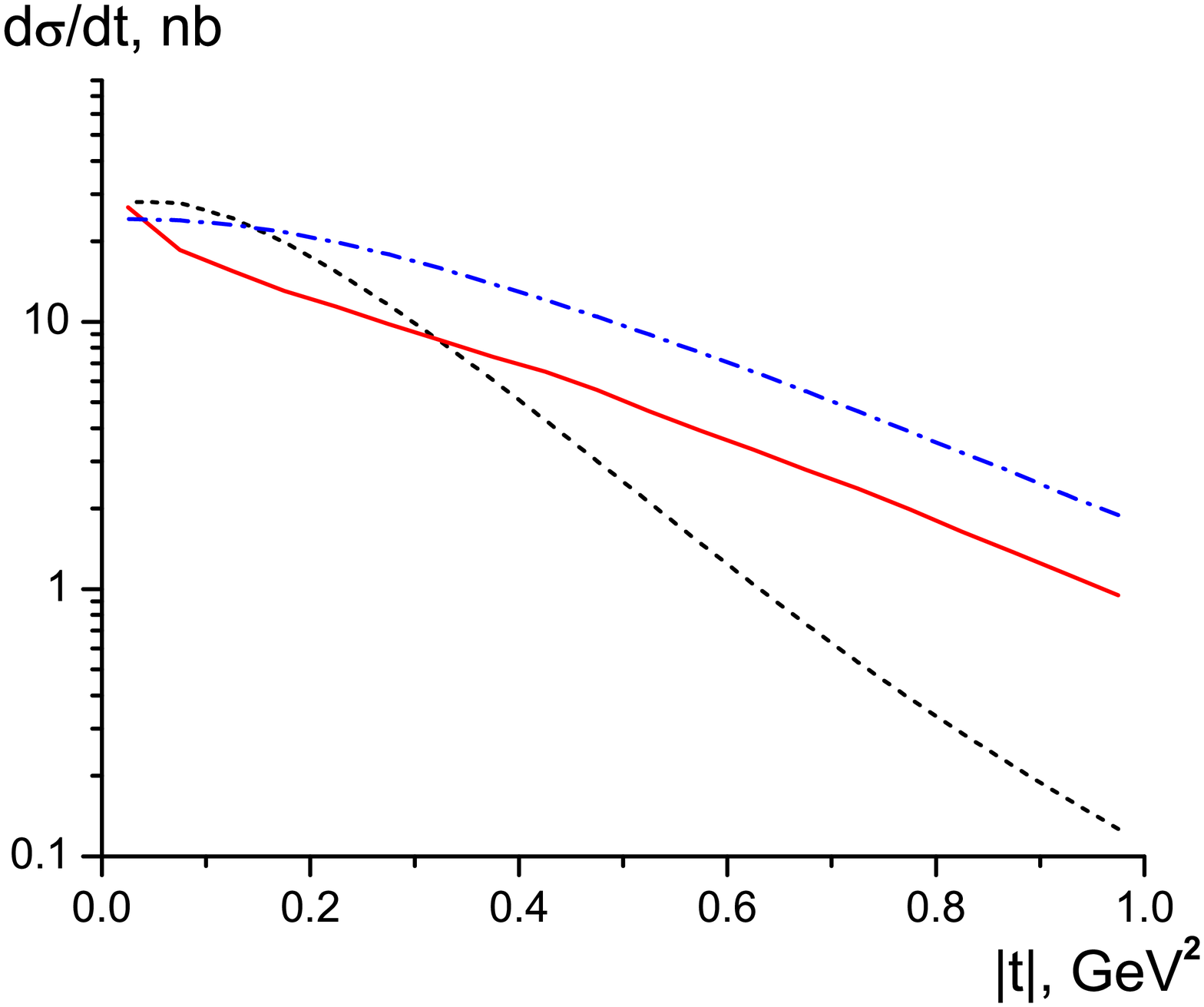}}
\end{minipage}
\begin{minipage}{0.32\textwidth}
 \centerline{\includegraphics[width=1.3\textwidth]{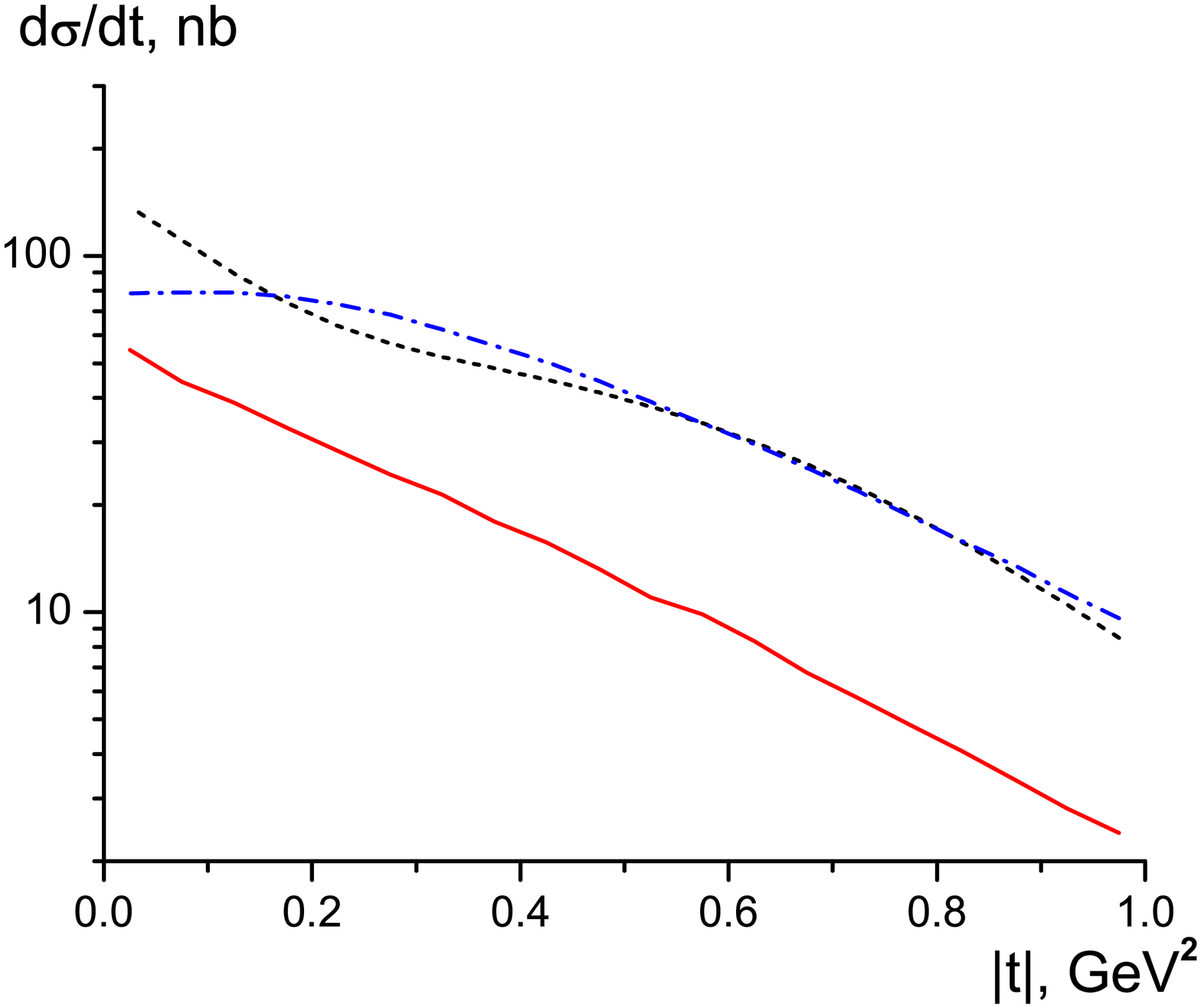}}
\end{minipage}
   \caption{\label{fig:chic012-dt}
   \small \em
Distribution in $t_{1,2}$ of $\chi_c(0^+)$ (left panel),
$\chi_c(1^+)$ (middle panel) and $\chi_c(2^+)$ (right panel) for meson
CEP for different UGDFs. The meaning of curves here is the same as
in Fig.~\ref{fig:chic012-dy}.}
\end{figure}

Finally, let us turn to differential distributions. In
Fig.~\ref{fig:chic012-dy} we show the differential cross section
$d\sigma/dy$ in rapidity $y$ for all $\chi_c$ states. In this figure
and in the following, all helicity contributions for
$\chi_c(1^+,2^+)$ CEP are taken into account. Here and below we show
only bare CEP cross sections for GBW, KS and KMR UGDFs. In the last
case, we present the results computed with the HKRS cut-off
parameter $0.72\,\GeV^2$ \cite{HarlandLang:2009qe}. We see that the
shape of the curves is rather similar, however, they have
substantially different maxima. The biggest cross section for the
$\chi_c(0^+,2^+)$ states is obtained with the KS UGDF, whereas for
$\chi_c(1^+)$ the KS and KMR UGDFs give quite similar cross
sections.

In Fig.~\ref{fig:chic012-dt} we present corresponding
distributions in $t = t_1$ or $t = t_2$ (identical), again for
different UGDFs. Except of normalisation the shapes are rather
similar. This is because of the $t_1$ and $t_2$ dependencies of form
factors (describing the off-diagonal effect) are taken the same for
different UGDFs.

\begin{figure}[!h]
\begin{minipage}{0.32\textwidth}
 \centerline{\includegraphics[width=1.3\textwidth]{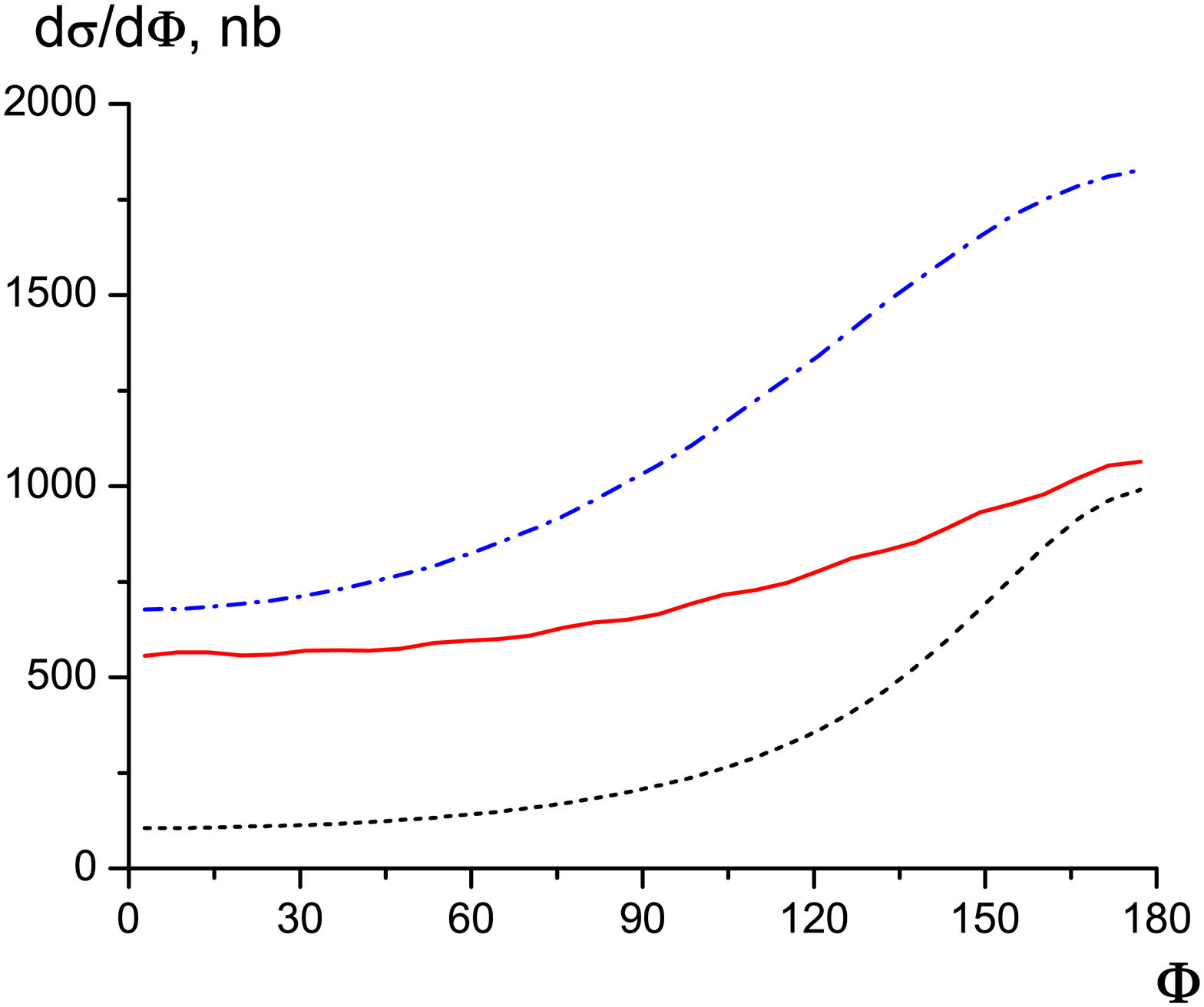}}
\end{minipage}
\begin{minipage}{0.32\textwidth}
 \centerline{\includegraphics[width=1.3\textwidth]{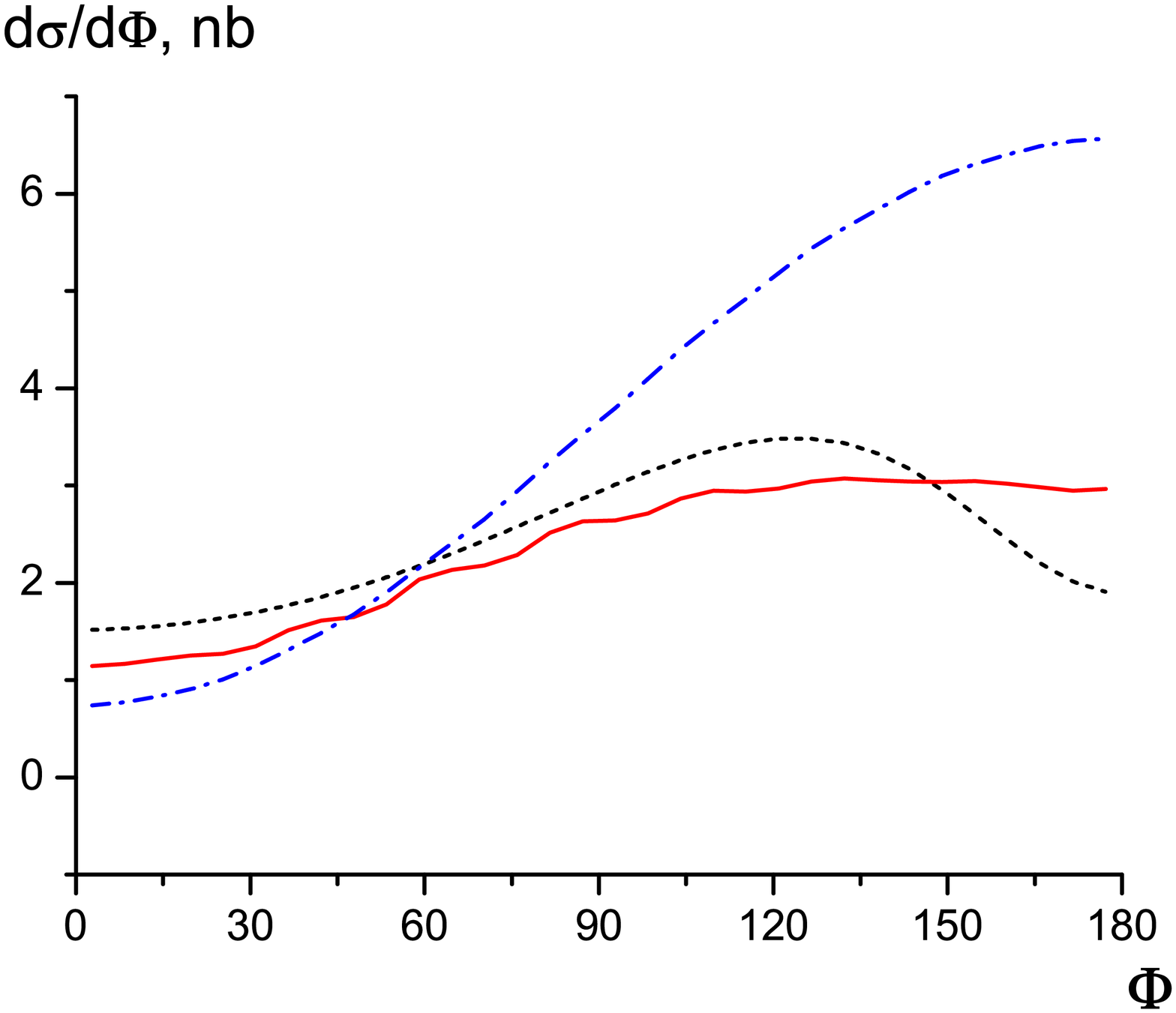}}
\end{minipage}
\begin{minipage}{0.32\textwidth}
 \centerline{\includegraphics[width=1.3\textwidth]{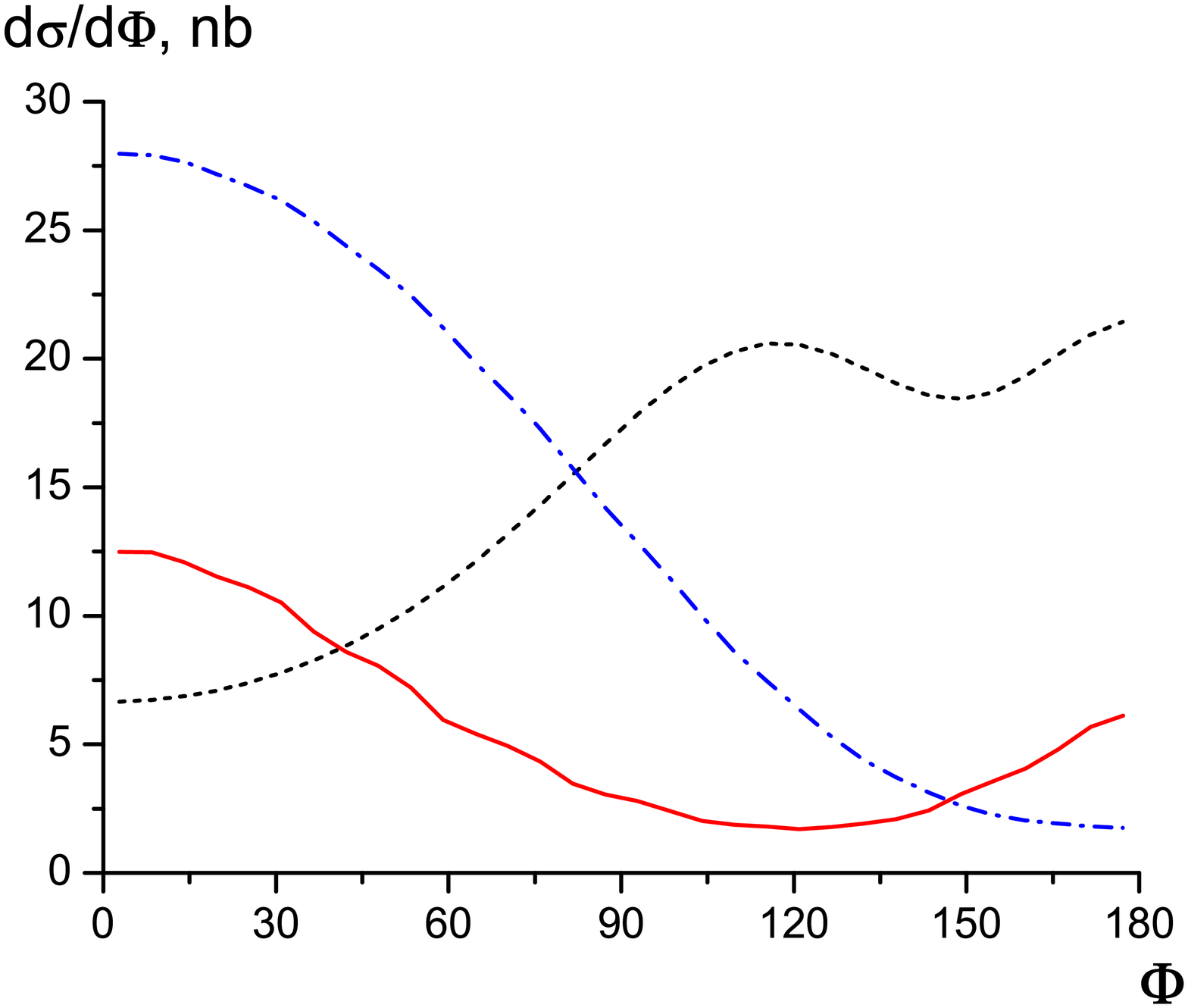}}
\end{minipage}
   \caption{\label{fig:chic012-dphi}
   \small \em Distribution in relative azimuthal angle $\Phi$
between outgoing protons for $\chi_c(0^+)$ (left panel),
$\chi_c(1^+)$ (middle panel) and $\chi_c(2^+)$ (right panel) meson
CEP for different UGDFs. The meaning of curves here is the same as
in Fig.~\ref{fig:chic012-dy}.}
\end{figure}

In Fig.~\ref{fig:chic012-dphi} we show the correlation function
$d\sigma/d\Phi$ in relative azimuthal angle $\Phi$ between outgoing protons
for different $\chi_c$ states. The shapes of the
distributions are somewhat dependent on UGDFs. It is interesting to
note here that the KS and KMR UGDFs lead to very similar angular
dependence of $d\sigma/d\Phi$ for all $\chi_c$ states. In the case
when energy resolution is not enough to separate contributions from
different states of $\chi_c$ ($\chi_c(0^+)$, $\chi_c(1^+)$,
$\chi_c(2^+)$), which seems to be the case for Tevatron, the
distribution in relative azimuthal angle may, at least in principle,
be helpful.

The fact that the angular distributions are not simple functions
(like $\sin\Phi$, $\cos\Phi$) of the relative azimuthal angle
between outgoing nucleons is due to the loop integral in
Eq.~(\ref{ampl}) which destroys the dependence one would obtain with
single fusion of well defined (spin, parity) objects (mesons or reggeons)
\cite{Pasechnik:2007hm}.

\section{Conclusions and discussion}

Our results can be summarized as follows:

We have derived the QCD amplitude for central exclusive production
of tensor $\chi_c(2^+)$ meson. This amplitude vanishes in the
forward limit of outgoing protons, as demanded by the $J_z=0$
selection rule. Our numerical results show the importance of
non-forward corrections, including all polarisation states of
$\chi_c(2^+)$ and nonperturbative contributions to the $\chi_c(2^+)$
CEP. Inclusion of all the ingredients leads to a noticeable
contribution of the $\chi_c(2^+)$ meson in the observable radiative
decay channel depending on UGDF. We have observed the importance of
the $\lambda=0$ state $\chi_c(1^+)$ CEP and $\lambda=0,\,\pm1$
states for $\chi_c(2^+)$ CEP at $y \approx$ 0, whereas the total CEP
cross section is dominated by maximal helicity contributions.

The main contribution to diffractive charmonium production comes
from small gluon transverse momenta $Q_t<1\,\GeV$ leading to quite
substantial sensitivity of the corresponding cross section on the
infrared cut-off in perturbatively modeled KMR UGDF. Alternatively
one could use UGDFs like Kutak-Sta\'sto and GBW models, which by
construction can be used for any values of the gluon transverse
momenta.

We have tested the symmetrical prescription for off-diagonal UGDFs,
following from positivity constraints and incorporating $x$, $q_t$
dependence of both participating gluons, against the present CDF
experimental data. A rather good quantitative agreement with the CDF
data on charmonium CEP in the radiative decay channel is achieved
with the nonlinear Kutak-Sta\'sto UGDF model giving the cross
section $d\sigma_{obs}(J/\psi\gamma)/dy(y=0)\simeq0.6$ nb without
imposing extra normalisation conditions beyond the QCD framework.
Such a description is achieved by incorporating very soft screening
gluons with $x'\sim 0.1\cdot q_{0,t}/\sqrt{s}$. We have also
calculated total cross sections of $\chi_c$ CEP at different
energies (RHIC, Tevatron and LHC), as well as differential
distributions in three phase space variables $y,\,t,\,\Phi$.

Overall theoretical uncertainty of the QCD mechanism under
consideration is rather high but hard to estimate due to large
unknown nonperturbative contributions coming into the game and not
well known higher-order QCD corrections to the hard subprocess
$g^*g^*\to\chi_c$ (especially, in the axial-vector case). Also,
absorptive corrections may depend on UGDF used in the calculation,
and there is no reliable estimation of such a sensitivity in
literature. In the present paper we kept the strategy to study
different distinct options and analyze the sensitivity of the final
results with respect to the UGDFs choice, prescriptions for skewed
UGDFs, nonperturbative cut-off parameter and characteristic $x'$
variations, etc. Then a comparison with experimental data would
allow to select the most reliable option. However, we observe a
variety of such ``good'' options, namely, description of the data
(with, however, pretty large theoretical uncertainties related, in
particular, with unknown NLO corrections) can be, in principle,
achieved for a few UGDFs (GBW, KS and KMR UGDFs, see
Table~\ref{table:dy0}). Each of them pick up some essential QCD
dynamics. Further constraints can, in principle, be settled by
experimental measurements of separate $\chi_c(J^+)$ contributions,
the energy dependence of the cross section and the shapes of
differential distributions.

\section{Acknowledgments}

Useful discussions and helpful correspondence with Mike Albrow,
Sergey Baranov, Rikard Enberg, W{\l}odek Guryn, Lucian Harland-Lang,
Gunnar Ingelman, Valery Khoze, Francesco Murgia, Mikhail Ryskin and
Wolfgang Sch\"afer are gratefully acknowledged. This study was partially
supported by the Polish grant of MNiSW N N202 249235, the Russian
Foundation for Fundamental Research, grants No. 07-02-91557, 08-02-00896, 09-02-00732  and No.
09-02-01149.

\end{document}